\documentclass{aa}  
\usepackage{lscape}
\usepackage{graphicx}
\usepackage{stfloats}
\usepackage[dvips]{color}
\usepackage{./bibtex/natbib}
\bibpunct{(}{)}{;}{a}{}{,}                           
\usepackage{txfonts}
\begin{document}
\bibliographystyle{./bibtex/aa.bst}
   \title{Core-collapse supernovae in low-metallicity environments and future all-sky transient surveys}

   \subtitle{}

   \author{D. R. Young \inst{1}\thanks{email: dyoung06@qub.ac.uk}
      \and S. J. Smartt \inst{1}
      \and S. Mattila \inst{1,2}
      \and N. R. Tanvir \inst{3}
      \and D. Bersier \inst{4}
      \and K. C. Chambers \inst{5}
      \and N. Kaiser \inst{5}
      \and J. L. Tonry \inst{5}}

   \institute{Astrophysics Research Centre, School of Maths and Physics, Queen's University Belfast, Belfast BT7 1NN, Northern Ireland, UK.
   \and Tuorla Observatory, University of Turku, V\"ais\"al\"antie 20, FI-21500 Piikki\"o, Finland.
   \and Department of Physics and Astronomy, University of Leicester, Leicester, LE1 7RH, UK.
   \and Astrophysics Research Institute, Liverpool John Moores University, Twelve Quays House, Egerton Wharf, Birkenhead, CH41 1LD, UK.
   \and Institute for Astronomy, University of Hawaii, 2680 Woodlawn Drive, Honolulu, HI 96822}

   \date{Received July 24, 2007; accepted July 24, 2007}
      
   \abstract
   {}
   {Massive stars in low-metallicity environments may produce exotic
explosions such as long-duration gamma-ray bursts and pair-instability
supernovae when they die as core-collapse supernovae (CCSNe). Such
events are predicted to be relatively common in the early Universe
during the first episodes of star-formation. To understand these
distant explosions it is vital to study nearby CCSNe arising in
low-metallicity environments to determine if the explosions have
different characteristics to those studied locally in high-metallicity
galaxies. Many of the nearby supernova searches concentrate
their efforts on high star-formation rate galaxies, hence biasing the
discoveries to metal rich regimes. Here we determine the feasibility of 
searching for these CCSNe in metal-poor dwarf galaxies 
using various survey strategies.} 
   {We determine oxygen abundances and star-formation rates for all
spectroscopically typed star-forming galaxies in the Sloan Digital Sky
Survey, Data Release 5, within $z=0.04$. We then estimate the CCSN
rates for sub-samples of galaxies with differing upper-metallicity
limits. Using Monte-Carlo simulations we then predict the fraction of
these CCSNe that we can expect to detect using different survey
strategies. We test survey capabilities using a single 2m telescope, a network 
of 2m telescopes, and the upcoming all-sky surveys of the Pan-STARRS
and LSST systems.} 
   {Using a single 2m telescope (with a standard CCD camera) 
search we predict a detection rate of
$\sim$1.3 CCSNe yr$^{-1}$ in galaxies with metallicities below
$12+\log({\rm O/H})<8.2$ which are within a volume that will allow
detailed follow-up with 4m and 8m telescopes ($z=0.04$).
With a
network of seven 2m telescopes we estimate $\sim$9.3 CCSNe yr$^{-1}$ 
could be found, although this would require more than 
1\,000\,hrs of telescope time allocated to the network. 
Within the same radial distance, a 
volume-limited search in the future Pan-STARRS
PS1 all-sky survey should uncover 12.5 CCSNe yr$^{-1}$  in low-metallicity
galaxies. Over a period of a few years this would
allow a detailed comparison of their properties. We then extend our 
calculations to determine the total numbers of CCSNe that can potentially be 
found in magnitude-limited surveys 
with PS1 (24\,000 yr$^{-1}$, within $z\lesssim0.6$), 
PS4 (69\,000 yr$^{-1}$, within $z\lesssim0.8$ ) 
and LSST (160\,000 yr$^{-1}$, within $z\lesssim0.9$) surveys.}
  {}
   
\keywords{supernovae -- gamma-rays bursts -- star-formation rate -- supernova rate}

\titlerunning{Core-collapse supernovae in low-metallicity environments}
\maketitle

\section{Introduction}

  Upon exhausting all of its available fuel, the iron core of a massive star ($\gtrsim$8$M_\odot$) collapses, often giving rise to an extremely powerful, bright explosion known as a core-collapse supernova (CCSN). Under very specific circumstances these CCSNe are also thought to be associated with another very powerful and luminous event known as a long-duration gamma-ray burst (LGRB). The circumstances that permit this extraordinary
partnership are still not well understood, but it seems likely
that progenitors born in lower-metallicity environments (sub-solar) favour production of GRBs \citep[see][]{2006AcA....56..333S,2006Natur.441..463F,2003A&A...406L..63F,2004MNRAS.352.1073T}.

  The most popular theoretical model interprets the LGRB as
  being caused by the core-collapse of a specific class of massive
  star. This fundamental theory, coined as the `Collapsar' model, was
  first conceived by \citet{1993ApJ...405..273W} \citep[also see][]{1999A&AS..138..499W}. The criteria for the creation of a LGRB
  resulting from the collapse of a massive star are that a black-hole
  must be formed upon core-collapse, either promptly or via fall-back
  of ejected material, and that an accretion disk is allowed to form
  around this black-hole which acts as a `central engine' causing the
  release of collimated relativistic jets along the rotation axis of
  the black-hole
  \citep{2006astro.ph.10276H,2008arXiv0801.4362Y,2008arXiv0803.3807T}. It
  is these relativistic jets that form the LGRB and the associated
  afterglow emission. These criteria are found to place 
  very specific
  constraints on the type of progenitor star that can collapse to
  produce a LGRB. The core of
  the progenitor must be massive enough to produce a black-hole, 
  it must be rotating rapidly enough to form an accretion disk, and the star
  needs to be stripped of
  its hydrogen envelope, so that the resulting relativistic jets are
  not inhibited from reaching the stellar surface. All of this points toward
  rapidly rotating Wolf-Rayet stars as viable progenitors. 
   \citep{2008arXiv0804.0014D,2006ApJ...637..914W,2006ARA&A..44..507W}

 Rapidly rotating O-stars which mix-up their core processed material
 to produce WR stars through chemically homogeneous evolution have been
suggested to 
  satisfy the requirements of the collapsar model \citep{1987A&A...178..159M,2006ASPC..353...63Y,2006A&A...460..199Y,2006ApJ...637..914W}.
 This model prevents the depletion of angular
 momentum due to mass-loss and allows a massive star to become a WR star
 directly, avoiding the supergiant phase. \citet{2006A&A...460..199Y} have estimated that for a
 rapidly rotating star to be the progenitor of an LGRB, prior to collapse
 it must have a helium core mass greater than 10M$_{\odot}$, that is an
 initial mass greater than 25-30M$_{\odot}$. \citet{2006ApJ...637..914W}
 have modelled the WR mass-loss rates of \citet{2005A&A...442..587V} and
 predict that the upper metallicity limit for forming LGRB may be as high
 as 0.3Z$_{\odot}$. Apart from the single star progenitor model, a rapidly rotating massive star could lose its hydrogen envelope
  and still retain a large fraction of its initial angular momentum if a close binary companion were to strip it of the envelope
  rapidly enough for the progenitor to avoid spin down
  \citep{2007A&A...465L..29C,2007astro.ph..2652B,2007A&A...471L..29D}.

Whatever the model for the progenitor of the LGRB, prior to collapse we require a very rapidly rotating, massive WR star. These constraints lead to the prediction that LGRBs favour a lower-metallicity environment and that CCSNe found in association with an LGRB will be classified as Type Ib or Type Ic, characterised by an absence of hydrogen in their spectra 
\citep{2007ApJ...666.1024B,2003fthp.conf..171F,2003astro.ph..1006H}. To date many LGRBs have revealed `bumps' in their afterglow emission, thought to be evidence of associated SNe \citep[see][]{1999Natur.401..453B, 2000ApJ...536..185G, 2003A&A...406L..33D,  2005ApJ...624..880L}. A few of these associated SNe have been adequately spectroscopically typed, all have been found to be broad-lined Type Ic SNe, helping to confirm the theory that LGRBs are the result of the collapse of hydrogen-stripped massive stars \citep{1998Natur.395..670G,2003ApJ...591L..17S,2003Natur.423..847H,2004ApJ...609..952Z}. \citet{2008AJ....135.1136M} have found that the SNe Type Ic-BL associated with an LGRB are found to inhabit a sample of host galaxies of significantly lower metallicity than the host galaxies of their Type Ic-BL counterparts with no associated observed LGRB. Interestingly, there has not been a SN Type Ic-BL without an associated observed LGRB that has been found to inhabit the same low-metallicity galaxy sample that the LGRB associated SNe inhabit.

\citet{2006Natur.441..463F} have found that LGRBs are preferentially
found in irregular galaxies which are 
significantly fainter, smaller and hence presumably of
lower-metallicity than typical CCSNe hosts. 
\citet{2005NewA...11..103S} make a similar point, stating that
the majority of LGRBs are found in extremely blue, sub-luminous
galaxies, similar to the Blue
Compact Dwarf Galaxy population.
\citet{2006astro.ph..9208S} 
has speculated that at primordial metallicities stars of up 
to 300M$_{\odot}$ may form and if they have low mass-loss, 
they may end their lives still retaining much of their original 
mass. \citet{2002ApJ...567..532H} predict that those stars
within the mass range 140M$_{\odot}$-260M$_{\odot}$ 
may produce  pair-instability supernovae
(PISNe), huge thermonuclear explosions with energies as high as
$\sim$10$^{53}$ ergs. Hence understanding SNe in low-metallicity 
environments locally could offer insights into high-z GRBs and
early Universe explosions. 
An indication that these exotic PISN
events may not be beyond our present day reach is the recent SN 2006gy
which has been reported to have a peak luminosity three times greater
than any other supernova ever recorded reaching a magnitude of -22
\citep{2006astro.ph.12617S}. Hypothesised as having had a Luminous
Blue Variable as a progenitor with an initial mass of
$\sim$150M$_{\odot}$ this has been suggested to be the first ever PISNe
detected \citep[also
see][]{2007arXiv0708.1970L,2007ApJ...659L..13O,2007Natur.450..390W}.
The unusual SN 2006jc defines another class of peculiar low-metallicity
event, occurring spatially coincident with an outburst thought to be
that of an LBV just two years previously \citep{2007Natur.447..829P}, 
which has also been postulated to be related to the PISN mechanism
\citep{2007Natur.450..390W}

To date there 
have been many surveys specifically designed to search for SNe and
depending on the motivation of an individual survey generally one of
two different survey strategies has been employed. The first of these
strategies is the \textit{pointed survey strategy}; here we have a
fixed catalogue of galaxies which are all individually
observed with a cadence of a few days and scanned for fresh SNe. These
surveys included the highly successful Lick Observatory Supernova
Search (LOSS) which uses the fully robotic 0.75m KAIT telescope to
scan a catalogue of 7\,500-14\,000 nearby galaxies ($z \lesssim 0.04$), 
aiming to observe each
individual galaxy with a cadence of 3-5 days
\citep{2001ASPC..246..121F}. With
the advent of relatively inexpensive CCD technology, the ability of
amateur astronomers to detect nearby SNe using this pointed survey
strategy has become more than substantial and a large fraction of
nearby SNe are now being discovered by this community
\citep{2006JAVSO..35...42G}.
The Southern inTermediate Redshift ESO
Supernova Search (STRESS) produced a catalogue of $\sim$43,000
($0.05 < z < 0.6$) 
galaxies within 21 fields observed with the ESO Wide-Field-Imager
and searched for SNe in these hosts to determine 
rates and their dependency on host galaxy colour and 
redshift evolution \citep{2008A&A...479...49B}.

The second strategy 
employed by SN surveys is the \textit{area survey strategy}; here an
area of sky is surveyed repeatedly and image subtraction used to
identify transient events including SNe. The SN Legacy Survey uses
the CFHT MegaCam imager to image four one-square degree fields to
search for SNe Type Ia with the motivation of improving the sampling
of these SNe within the redshift range $0.2 < z < 1.0$
\citep{2005ASPC..339...60P, 2006A&A...447...31A}. The Equation of
State SupErNova Cosmology Experiment (ESSENCE) uses the Mosaic Imager
on the Blanco 4m telescope to survey equatorial fields, that have
preferably been observed with other wide-field surveys, to discover SN
Type Ia within the redshift range $0.15 < z < 0.74$
\citep{2007ApJ...666..674M}.  The Nearby SN Factory uses ultra-wide
field CCD mosaic images from the Near-Earth Asteriod Tracking (NEAT) and Palomar-Quest survey
programs to perform an area survey in a bid to find SN Type
Ia in the redshift range $0.03 < z < 0.08$
\citep{2002SPIE.4836...61A}. The SDSS-II Supernova Survey takes
repeated images of Stripe 82, a 300 square degree southern equatorial
strip of the sky, driven by the ambition of identifying and measuring
the lightcurves of SNe Type Ia in the intermediate redshift range
$0.05 < z < 0.35$ \citep{2008AJ....135..338F, 2008AJ....135..348S}. 
The Texas Supernova Search search 
\citep{2005AAS...20717102Q, 2007AAS...21110505Y}
is somewhat unique in that it 
has a small aperture telescope (0.45m) but very wide field (1.85 square
degrees) and is focussed on finding nearby SNe in wide area searches. 
This survey discovered SN2006gy along with many other of the brightest
known SNe ever found and we shall discuss this further in 
Sec. \ref{sect:disc}.

The main driver of 
SN surveys that employ the pointed survey strategy is basically to
find as many SNe as possible regardless of type or characteristic.  To
ensure that these surveys are as efficient as possible at finding
CCSNe, the galaxies catalogues used by the surveys generally consist
only of the most massive galaxies with the greatest star-formation
rates. As a result these galaxy catalogues
tend to be heavily biased towards galaxies with high metallicity. This
over arching high-metallicity bias has more than likely placed us in
the situation where the vast majority of CCSNe that have occurred in
low-metallicity environments (such as low-luminosity, dwarf, irregular
galaxies) have remained undetected.  \citet{2008ApJ...673..999P} have
matched the SAI supernova catalogue to the SDSS-DR4 catalogue of
star-forming galaxies with measured metallicities and it is clear that
the vast majority of the SNe considered have been detected in areas of
relatively high-metallicity; SN Type II occurring in host galaxies
with mean metallicity \mbox{$12+{\rm \log(O/H)}=8.94\pm0.04$} and SN
Ib/c at 9.06$\pm$0.04.

It is inherently interesting then to search for SNe specifically in 
low-metallicty environments, or at least without biasing the search
to look at only high metallicity regimes. 
In this paper we shall discuss several methods to 
search for low-metallicity CCSN
events. First we consider compiling a catalogue of nearby,
low-metallicity galaxies from pre-existing catalogues 
(taking SDSS DR5 as our primary survey source)
and using either a single 2.0m telescope or a
network of 2.0m telescopes to perform a pointed survey of 
low-metallicity galaxies in the hope of detecting a few CCSNe. 
Secondly we consider
using a future all-sky transient survey such as the Panoramic Survey
Telescope and Rapid Response System (Pan-STARRS) to perform a
volume-limited survey for SNe and estimate how many may be found
in low-metallicity galaxies. 
A third and final method we consider is to use a future all-sky
transient survey, limited only by
the limiting magnitude of the survey, to search for all CCSNe
including those low-metallicity events.While this paper is 
primarily aimed at determining numbers of low metallicity events
that could be found, the latter calculation gives an estimate of 
the total number of CCSNe that are likely to be harvested from these
upcoming surveys.

\section{Creating Galaxy Catalogues for the Various Survey Strategies}\label{sec_a}

  In order to produce catalogues of galaxies that we can use for the various survey strategies to be considered, we begin with the Sloan Digital Sky Survey (SDSS) \citep{2007arXiv0707.3380A}. The SDSS Data
  Release 5 (DR5) provides photometric
  and spectroscopic data, within a wavelength range of 3500-9200$\AA$, for
  $\sim$675\,000 galaxies, over an area of 5\,713 square degrees of the northern
  hemisphere out to a redshift
  of $z\sim0.4$. In general we only want to detect relatively nearby CCSNe as we want to spectroscopically type and follow these SNe with relative ease, so to this end we introduce a distance limit of  $z=0.04$ to the SDSS spectroscopic catalogue. We have used
  the SDSS DR5 website\footnote{SDSS DR5 website:
  \it{http://www.sdss.org/dr5}} to extract the out 44\,041 galaxies within z
  $=0.04$ along with data
  including the petrosian magnitudes ({\it
  u, g, r, i} and  {\it z}), the galactic extinctions in each filter determined following \citet{1998ApJ...500..525S}, the spectroscopic
  redshifts, the
  {\it r}-band fibre magnitudes and the line intensities of H$\alpha$, H$\beta$,
  [OIII]$\lambda5007$ and [NII]$\lambda6584$ for each galaxy.
  
 Of these 44\,041 galaxies we extract out two separate samples of galaxies. The first of these samples is classified as the high signal-to-noise sample, containing galaxies with an SDSS defined line significance indicator `nSigma'  $>$ 7$\sigma$ for all of the four spectral lines mentioned previously (as advised by DR5). The second sample is classified as the low signal-to-noise sample, containing galaxies that have not been included in the high signal-to-noise sample but exhibit `nSigma' $>$ 5$\sigma$ in both H$\alpha$ and H$\beta$. The high signal-to-noise sample contains 20\,632 galaxies and the low signal-to-noise sample contains 8\,703 galaxies.
  
\subsection{\textbf{Removing AGN}}\label{sec_aa}

  Eventually we aim to determine a star-formation rate (SFR) for each
  individual galaxy
  in our sample in order to then determine their core-collapse SN rates
  (CCSR).
  As it is the young, massive, hot star population within each galaxy that
  is the dominant source of hydrogen ionising radiation, it is possible to use
  the H$\alpha$ luminosity of these galaxies as a SFR indicator.
  
  A difficultly arises in that many galaxies within both the high signal-to-noise and the low signal-to-noise galaxy samples will host Active Galactic Nuclei
  (AGN), which will also contribute to the galaxy's H$\alpha$
  luminosity. To remove these AGN contaminated galaxies
  from the high signal-to-noise galaxy sample we use the following
  diagnostic line provided by \cite{2003MNRAS.346.1055K} to discriminate between
  purely star-forming galaxies (SFGs) and galaxies that also host AGN:
  
\begin{equation}
  \log\left(\frac{[\rm{OIII}]\lambda5007}{\rm{H}\beta}\right)=\frac{0.61}{\log([\rm{NII}]\lambda6584/\rm{H}\alpha)-0.05}+1.3
  \label{kauff_equ}
\end{equation}

\begin{figure}[t]
  \begin{center}
  \includegraphics[totalheight=0.2\textheight]{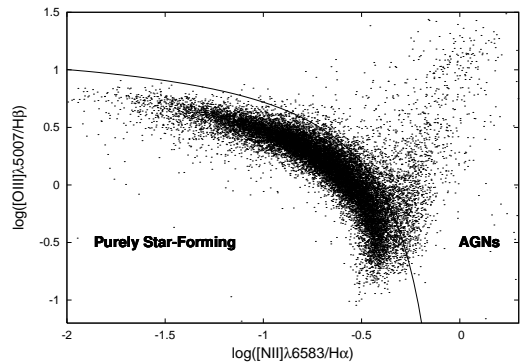}
  \caption{\textrm{[OIII]$\lambda5007/$H$\beta$ vs
  [NII]$\lambda6584/$H$\alpha$ plot of 20\,632 galaxies in our SDSS DR5 good signal-to-noise galaxy sample.
  The diagnostic line from \cite{2003MNRAS.346.1055K} discriminates between the purely star-forming galaxies in
  our sample (those below the line) and those hosting AGN (above the line).}\label{BPT_dia}}
  \end{center}
\end{figure}

  \noindent Fig. \ref{BPT_dia} shows the SFGs found below this line and AGN
  host galaxies above the line. We now have 18\,350 high signal-to-noise SFGs. Concerning the low signal-to-noise galaxy sample , we do not have accurate enough spectral information to apply this diagnostic line to remove any unwanted AGN host galaxies. However, following the example of
  \citet{2004MNRAS.351.1151B} it is still possible to remove the AGN hosts from
  the low signal-to-noise sample by requiring that [NII]$\lambda6584$/H$\alpha > 0.6$ and that
  nSigma $>$ 7$\sigma$ in both lines. We now also have 6\,000 low signal-to-noise SFGs. An overview of all galaxy sub-samples can be found
  in Table \ref{subsample_table}. The 17\,409 unclassified galaxies are
  predominantly early-type galaxies that show little signs of
  recent star-formation, which is the reason that H$\alpha$ has not been detected to
  the significance levels that we have required. The lack of recent star-formation within these galaxies implies that they will
  also be void of any future CCSN events and hence these galaxies are not of
  interest to our survey.
  
  Having removed all of the AGN contaminated galaxies we now have two catalogues of SFGs which we shall refer to from now on as the high signal-to-noise SFG (HSFG) catalogue and the low signal-to-noise SFG (LSFG) catalogue.

\begin{table*}
  \begin{center}
  \caption{\textrm{Hierarchy of galaxies extracted from the original SDSS
    DR5 spectroscopic galaxy sample within z$=$0.04}\label{subsample_table}}
    \begin{tabular}{c | c | c | c | c | c | c}
      \hline\hline
      \multicolumn{7}{c}{\bf All SDSS DR5 Galaxies [44\,041]}\\
      {\bf Sample} & \multicolumn{2}{|c|}{High signal-to-noise galaxies [20\,632]} & \multicolumn{2}{|c|}{Low signal-to-noise galaxies [8\,703]}& \multicolumn{2}{c}{Unclassified galaxies [17\,409]}\\
      {\bf Sub-sample} & {SFGs [18\,350]} & {AGN [2\,282]} & {SFGs [6\,000]} & {AGN [2\,703]} & \multicolumn{2}{c}{}\\
      \hline
    \end{tabular}
  \end{center}
\end{table*}

\subsection{Measuring Oxygen Abundances}\label{sec_b}

  In order to select out low-metallicity galaxies from both the HSFG
  and LSFG catalogues,
  we define oxygen abundances using the empirical calibrations of
  \citet{2004MNRAS.348L..59P}. Within the range $8.12 \lesssim 12+\log(\rm{O/H}) < 9.05$ the
  following empirical calibration is used:

\begin{center}  
\begin{equation}
  12+\log(\rm{O/H})=8.73-0.32\log\left(\frac{[\rm{OIII}]\lambda5007/\rm{H}\beta}{[\rm{NII}]\lambda6584/\rm{H}\alpha}\right),
\end{equation}
\end{center}

  \noindent and below $12+\log(\rm{O/H})\simeq 8.12$ the following calibration is used:

\begin{center}  
  \begin{equation}
    12+\log(\rm{O/H})=8.9+0.59\log([\rm{NII}]\lambda6584/\rm{H}\alpha).
  \end{equation}
\end{center}

\begin{figure}
  \begin{center}
  \includegraphics[totalheight=0.2\textheight]{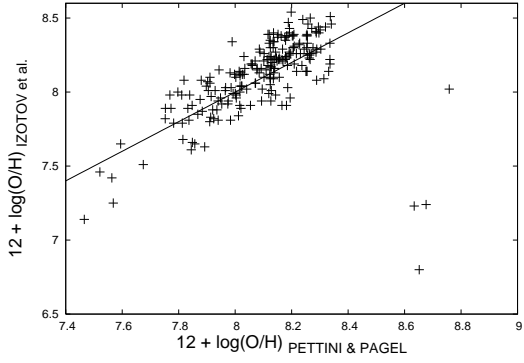}
  \caption{\textrm{Comparing our \citet{2004MNRAS.348L..59P} empirically calibrated oxygen abundances with those
  directly measured by \citet{2006A&A...448..955I} via measuring
  the electron temperature reveals that our
  measurements are reliable with an RMS scatter of 0.14 dex, the solid line depicting a one-to-one correspondence. Note the four outlying galaxies - see text for
  details.}\label{izo_fig}}
  \end{center}
\end{figure}

  The wavelengths of the emission lines used for the flux ratios in both of these calibrations
  are separated by only a small amount and therefore their ratios are free of any extinction
  effects. Of the sample of 18\,350 HSFGs, \citet{2006A&A...448..955I} have directly measured
  metallicities for 209. They determined the
  oxygen abundances by measuring the
  [OIII]$\lambda$4363/[OIII]$\lambda$5007 line ratio to calculate an
  electron temperature $T_{e}$, and then derived the
  abundances directly from the strengths of the [OII]$\lambda$3727 (or
  [OII]$\lambda$7320, 7331 when [OII]$\lambda$3727 was not available) and
  [OIII]$\lambda\lambda$4959, 5007 emission lines. Comparing the
  \citeauthor{2006A&A...448..955I} $T_{e}$ measured abundances with our  \citeauthor{2004MNRAS.348L..59P}
  empirically calibrated abundances we find good agreement, apart
  from four outlying galaxies (see Fig. \ref{izo_fig}). These galaxies are
  SDSS J124813.65-031958.2, SDSS J091731.22+415936.8, SDSS
  J123139.98+035631.4 and SDSS J130240.78+010426.8.

  When viewed, these four
  outliers seem to be dwarf galaxies that are in the same
  line of sight as and possibly gravitationally bound
  to much larger and presumably more metal rich galaxies.
  Assuming that the contamination from these background galaxies has not been
  adequately removed from the spectra of the dwarf galaxies by the SDSS reduction pipeline would explain
  the discrepancy between the oxygen abundances measured by \citeauthor{2006A&A...448..955I} and
  our measurements. Viewing a random selection of the remaining galaxies from
  the \citeauthor{2006A&A...448..955I} sample reveal the blue compact dwarf galaxies expected from
  their low-metallicity galaxy sample.
  The fact that we may be over-estimating the oxygen abundances of a very
  small fraction of galaxies should not concern us too much as we are trying to produce a
  low-metallicity galaxy sample and not a high-metallicity sample that would then
  possibly be contaminated by a few misplaced lower-metallicity galaxies. Choosing then
  to ignore these four outlying galaxies we
  find that our oxygen abundance measurements for the remaining
  205 galaxies fall with an RMS scatter of 0.14 dex from the directly
  measured abundances of the \citeauthor{2006A&A...448..955I}
  
 Recently \cite{2008ApJ...673..999P} have taken a sample of 125 958 SFGs from SDSS DR4, with oxygen abundances derived in the same fashion as  \cite{2004ApJ...613..898T} used for SFGs in DR2. The \cite{2004ApJ...613..898T} method for deriving oxygen abundance estimates an individual galaxy's metallicty via a likelihood analysis which simultaneously fits multiple optical nebular emission lines to those predicted by the hybrid stellar-population plus photoionisation models of \cite{2001MNRAS.323..887C}. A likelihood distribution of the metallicity is determined for each galaxy the median is taken as the best estimate of the galaxy metallicity. The \cite{2004ApJ...613..898T} metallicities are essentially on the \cite{2002ApJS..142...35K} abundance scale. Matching our catalogues of HSFGs and LSFGs against the sample of 125 958 SFGs of \cite{2008ApJ...673..999P}, we find a common sample of 18 014 SFGs. The oxygen abundances that we measure with the \citeauthor{2004MNRAS.348L..59P} method are typically $\sim$0.2 dex below that of \citeauthor{2004ApJ...613..898T}, in agreement with the findings of \cite{2008AJ....135.1136M}. The cause of this discrepancy is debated but may either be due to certain parameters that produce temperature variations not being taken into consideration when deriving $T_{e}$ at higher-metallicity, which would lead to an under-estimation of the oxygen abundance measured on the \citeauthor{2004MNRAS.348L..59P} scale which is calibrated with $T_{e}$ measured abundances \citep{2005A&A...434..507S, 2006astro.ph..8410B}, or to an unknown problem with the photoionisation models used by  \citeauthor{2004ApJ...613..898T} \citep{2003ApJ...591..801K}. 
  
\subsection{SN Rate Indicator}\label{sec_c}

  Having measured the metallicity for each of the galaxies in our
 catalogues, we now wish to determine CCSRs. To do this we must first
  determine SFRs for the galaxies and then determine the fraction of those stars
  formed that will eventually end their lives as CCSNe. The
  best indicator that we have for the SFR for each galaxy is its H$\alpha$
  luminosity. As alluded to previously, it is going to be the young, massive, hot stars
  in purely star-forming galaxies that are the dominant source of
  hydrogen ionising radiation, causing the galaxy's
  H$\alpha$ luminosity to be proportional to its recent SFR.
  
  \citet{1998ARA&A..36..189K} has determined the following calibration
  between a galaxy's SFR and its H$\alpha$ luminosity:
  
\begin{center}
\begin{equation}
  {\rm SFR_{H\alpha}}({\rm M}_{\odot}{\rm \,\,yr^{-1}})=\frac{L_{{\rm H}\alpha}}{1.27\times10^{34}({\rm W})}
  \label{kennicutt_equ}
\end{equation}
\end{center}

  \noindent where the luminosity is measured in Watts.

  Derived from their model fits, \citet{2004MNRAS.351.1151B} also provide likelihood distributions for
  the conversion factor between the H$\alpha$ luminosity and the SFR for galaxies of
  various mass ranges. They confirm that the Kennicutt calibration is a
  very good {\it typical} calibration, comparing well with
  the median value for their sample. When considering the complete HSFG and LSFG catalogues it is acceptable to assume a median mass range for the galaxies and we employ
  the Kennicutt calibration. However, when considering galaxies with
  relatively poor metallicity we choose to use the most
  probable conversion factor for the \citeauthor{2004MNRAS.351.1151B} distribution with the lowest mass range (${\rm \log}M_{*}<8$) as this
  distribution closest resembles the low-metallicity galaxies in our catalogues i.e. low-mass, blue,
  dwarf, irregular galaxies.

\begin{center}  
\begin{equation}
  {\rm SFR_{H\alpha}}({\rm M}_{\odot}{\rm \,\,yr^{-1}})=\frac{L_{{\rm H}\alpha}}{2.01\times10^{34}({\rm W})}
\end{equation}
\end{center}

\begin{figure*}[b]
\begin{center}  
\begin{equation}
  {\rm SFR_{H\alpha}}({\rm M}_{\odot}{\rm \,\,yr^{-1}})=10^{-0.4(r_{\rm Petro} - r_{\rm
  fibre})}\left[\frac{S_{\rm H\alpha}/S_{\rm H\beta}}{2.86}\right]^{2.114}\frac{L_{{\rm H}\alpha}}{1.27\times10^{34}({\rm W})}
   \label{AC_equ}
\end{equation}
\end{center}
\end{figure*}

\begin{figure*}[b]
\begin{center}  
\begin{equation}
  {\rm SFR_{H\alpha}}({\rm M}_{\odot}{\rm \,\,yr^{-1}})=10^{-0.4(r_{\rm Petro} - r_{\rm
  fibre})}\left[\frac{S_{\rm H\alpha}/S_{\rm H\beta}}{2.86}\right]^{2.114}\frac{L_{{\rm H}\alpha}}{2.01\times10^{34}({\rm
  W})}
   \label{AC_low_equ}
\end{equation}
\end{center}
\end{figure*}

\begin{figure}
  \begin{center}
  \includegraphics[totalheight=0.2\textheight]{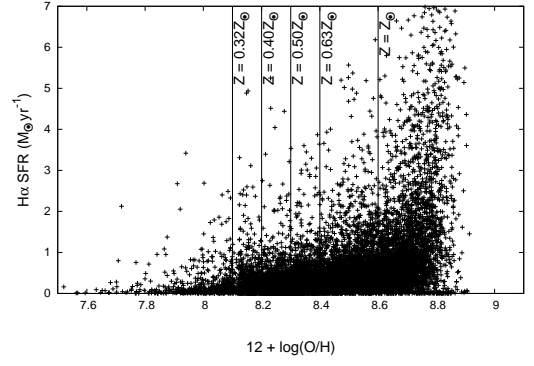}
  \caption{\textrm{H{$\alpha$} determined star-formation rates using the low-mass range calibration of \citet{2004MNRAS.351.1151B} of
  18\,350 galaxies
  in our HSFG catalogue compared with their measured oxygen abundances. In general,
  the greater the galaxy metallicity, the greater its star-formation rate - as
  expected.}
\label{oxy_sfr}}
  \end{center}
\end{figure}

  SDSS provides the H$\alpha$ equivalent line width and also the continuum flux
  at the wavelength of H$\alpha$, the equivalent width times the continuum flux
  giving the H$\alpha$ flux which is then used to determine the H$\alpha$
  luminosity. A problem with the SDSS data is that the measured H$\alpha$
  flux for a galaxy is only the flux which falls within the 3''
  fibre aperture of the SDSS multi-fibre spectrograph. Typically this is only a
  fraction of the total galaxy flux, as the SDSS spectrograph fibre locks onto the
  centre of the galaxy and any flux that falls outside of the 3'' fibre
  aperture is lost. \citet{2003ApJ...599..971H} have developed a very simple aperture-correction that can
  be applied to the measured galaxy H$\alpha$ luminosity to give an estimate of the total
  galaxy  H$\alpha$ luminosity. The aperture-correction takes account of the ratio between
  the petrosian photometric $r$-band galaxy magnitude and the
  synthetic $r$-band `fibre magnitude' determined from the galaxy spectrum. 
  \citeauthor{2003ApJ...599..971H} also provide an extinction correction to be
  used when determining the galaxy H$\alpha$ SFR. The correction makes use of the Balmer decrement
  and assumes the standard Galactic extinction law of
  \citet{1989ApJ...345..245C}. This gives
  a final aperture and extinction corrected H$\alpha$ luminosity SFR indicator
  for our entire galaxy sample as given in Equation \ref{AC_equ}, and for the low-metallicity galaxy sample as given in Equation \ref{AC_low_equ}, where $S_{\rm H\alpha}$ and $S_{\rm H\beta}$ are the line 
fluxes corrected for stellar absorption according to \citet{2003ApJ...599..971H}.
   Having now determined a SFR for each of the galaxies in our
  catalogues, we can compare these rates with their oxygen abundances
  (Fig. \ref{oxy_sfr}). It is
  clear that in general the higher the galaxy SFR the higher the typical oxygen abundance. It can be reasoned that if the SFR of a galaxy is high, or has been high at any point in its history, there is an increased population of young, hot, massive stars which in turn leads to an increased rate of
  CCSNe and therefore a greater rate of enrichment of the
  ISM. The observed high-metallicity cutoff of the SDSS galaxies is probably due to a saturation suffered by [OIII] $\lambda$4363 T$_{e}$ calibrated metallicities \citep[see][]{2008arXiv0801.1849K}

  Given a SFR, it is relatively simple to convert to a CCSR by determining
  the fraction of a stellar population that will eventually collapse to create
  CCSNe. Following the method used by e.g. \citet{2001MNRAS.324..325M}, we use an initial mass function (IMF) for a stellar
  population to calculate the fraction of this population within the mass
  range 8M$_{\odot}<$ M $<$ 50M$_{\odot}$, the same mass range for stars predicted to end
  their lives as CCSNe. From this logic the CCSR is determined as:

\begin{center}  
\begin{equation} 
  {\rm CCSR}=\frac{\int_{\rm 8M_{\odot}}^{\rm 50M_{\odot}}\phi(m)dm}{\int_{\rm
  0.1M_{\odot}}^{\rm 125M_{\odot}}m\phi(m)dm}\times{\rm SFR}
  \label{mattila_equ}
\end{equation}
\end{center}

  where \mbox{$\phi(m)$} is the Salpeter IMF \citep{1955ApJ...121..161S} with upper and
  lower mass cut-offs of \mbox{$0.1{\rm M}_{\odot}$} and \mbox{$125{\rm M}_{\odot}$}. This conversion is
  calculated to be:

\begin{center}  
\begin{equation} 
  {\rm CCSR (SNe \,\,yr^{-1})}=0.007 \times {\rm SFR (M_{\odot}\,\,yr^{-1})}
  \label{mattila_equ}
\end{equation}
\end{center}

\subsection{Additional nearby bright galaxies}

  The target selection criteria for the SDSS DR5 spectroscopic sample
  includes a
  bright magnitude limit of \mbox{$r\sim$14.5} in order to avoid saturation and excessive
  cross-talk in the spectrographs. As a result of this restriction many of
  the nearby luminous galaxies have been omitted from the DR5 spectroscopic
  sample. To account for these missing bright galaxies and 
construct a complete  galaxy catalogue we initially
  select out the galaxies from the HyperLeda galaxy catalogue that match with
  galaxies from the SDSS
  DR5 {\it photometric} survey with magnitudes $r<14.5$, assuming that the HyperLeda
  catalogue contains all of the nearby luminous galaxies. From this catalogue
  of matched galaxies we further select out those galaxies that are not found in the
  SDSS DR5 {\it spectroscopic} survey. We discover a total of 1\,887 nearby
  luminous galaxies included in the SDSS DR5 photometric survey that have
  been omited from the spectroscopic survey.
  
  Of these 1\,887 galaxies we wish to know the fraction that are star-forming
  galaxies. HyperLeda provides a galaxy morphological classification for a
  large number of the galaxies in the catalogue and we are able to remove all
  galaxies with an early-type classification, as these galaxies
  will generally have very low SFRs. But note that some early-type galaxies
  have HII regions and some evidence of low-level star-formation
  \citep[e.g.][]{2005MNRAS.357.1337M}. The mean
  $g-i$ colour of these removed early-type galaxies is 1.216, with a standard
  deviation \mbox{$\sigma = 0.153$}. In an attempt to remove the remaining
  non-starforming galaxies from this bright galaxy sample that do not have a morphological classification, we remove all
  galaxies redward of one $\sigma$ from the mean $g-i$ colour, i.e.
  $g-i>1.063$. This results in 1\,216 remaining star-forming galaxies. There is a possibility that this cut may also exclude a few starburst galaxies which have been heavily reddened due to the effects of extinction, but we are mainly concerned about low-metallicity galaxies, which are not greatly affected by extinction.
  
  As we have no spectral information for these galaxies from SDSS DR5, it is
  necessary to use an alternative indicator to estimate SFR other than the
  H$\alpha$ luminosity. The $U$-band luminosity can be used as a suitable
  indicator as developed by \cite{2006ApJ...642..775M}:
  
\begin{center}  
\begin{equation}
  {\rm SFR_{\it U_{obs}}}({\rm M}_{\odot}{\rm yr^{-1}})=(1.4\pm1.1)\times10^{-43}\frac{L_{U_{obs}}}{\rm ergs\ s^{-1}},
\end{equation}
\end{center}

  \noindent where the SDSS $u$-band is transformed to \mbox{$U_{\rm vega}$} using the following
  transformation from \citet{2007AJ....133..734B}:
  
\begin{center}  
\begin{equation}
  U_{\rm vega}=u-0.0140(u-g)+0.0556,
\end{equation}
\end{center}
 
 \noindent and the distance moduli from HyperLeda are used to determine $L_{U_{obs}}$. The
 CCSR is then determined using Equation \ref{mattila_equ}. Note
 that as \citeauthor{2006ApJ...642..775M} have empirically calibrated this
 $U$-band SFR indicator from
 extinction-corrected H$\alpha$ galaxy luminosities, there is no need to
 further correct this indicator for dust reddening. The 1\,216 
 bright, nearby galaxies have a resulting $U$-band indicated CCSR of \mbox{12.70
 CCSNe yr$^{-1}$}.

\begin{figure}
  \begin{center}
  \includegraphics[totalheight=0.3\textheight]{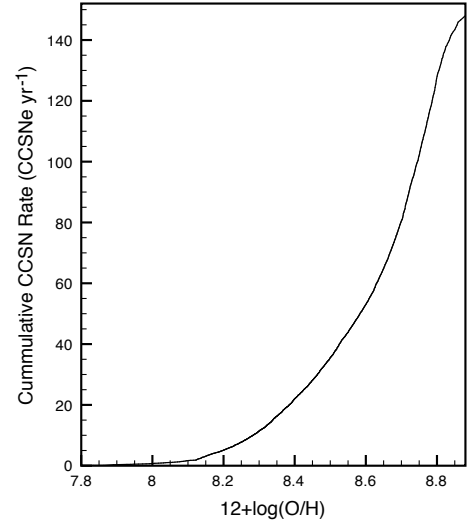}
  \caption{\textrm{Cumulative distribution of the combined CCSN rate from our HSFG and LSFG catalogues, measured against oxygen abundance. It is clear that as oxygen abundance increases so to does the rate of CCSNe}}\label{cum_dis}
  \end{center}
\end{figure}

  We now have three galaxy catalogues; a HSFG catalogue and a LSFG
  catalogue, with measured metallcities and CCSRs, and a nearby 
  galaxy catalogue containing bright galaxies not found in the SDSS
  spectroscopic galaxy catalogue. 
  We have estimates of SFRs and CCSRs for all these galaxies.
  Taking our
  catalogues of 18\,350 HSFGs and 6\,000 LSFGs and introducing various
  upper limits on the individual galaxy metallicities and lower limits
  on the CCSRs we are able to extract out separate galaxy
  sub-samples. Table \ref{SFG_rate_table} displays these sub-samples,
  differing both in the number of galaxies they contain and their
  estimated CCSR determined from the galaxy SFRs
    derived from aperture and extinction-corrected H$\alpha$ luminosities for the HSFG and LSFG catalogues, and from $U$-band luminosities for the 
nearby bright galaxy catalogue. For the samples of
    galaxies with no metallicity constraint the SFRs have been derived using the
    `typical' calibration of \citet{1998ARA&A..36..189K} whereas the SFRs of the
    sub-samples of galaxies with constraints on metallicity have been
    derived using the lowest mass range calibration of \citet{2004MNRAS.351.1151B}. These full catalogues or sub-samples of these catalogue can now be used to determine the feasibility of searching for low-metallicity CCSN events using various survey strategies. Figure \ref{cum_dis} shows the cumulative
  distribution of the combined CCSN rate from our HSFG and LSFG
  catalogues, measured against oxygen abundance. It is clear that the
  CCSR increases steeply with oxygen abundance of the galaxy sample.

\section{\textbf{Strategy 1 : A Pointed survey of catalogued low-Metallicity galaxies}}\label{sec_d}

  When determining the feasibility of using a pointed low-metallicity survey to search for CCSNe in a catalogue of low-metallicity galaxies we consider using the fully robotic 2.0m
  Liverpool Telescope situated at the Observatorio del Roque de los Muchachos, La Palma. We shall also consider the using of a network of similar sized telescopes.
  
  There are three characteristics that we require of the galaxy catalogue that we shall use with this survey strategy; firstly, it must contain only a few hundred galaxies as too many galaxies in the catalogue would hinder our ability to observe each individual
  galaxy frequently enough to ensure that we detect any SN that it may
  host. Secondly, we require that the galaxies are of
  sufficiently low metallicity in order to determine how the CCSNe that
  they host differ from those hosted by their higher metallicity counterparts.
  Finally, the galaxies must exhibit a suitably high enough CCSR to increase
  the probability of detecting these CCSNe. The latter two requirements somewhat
  contradict each other because metallicity
  tends to scale with SFR in SFGs and therefore requiring a
  low-metallicity
  galaxy sample implies that we require galaxies with lower
  SFRs, and hence {\it lower} CCSRs. It is
  therefore essential that we produce a galaxy catalogue that can act as
  a compromise between these two conflicting requirements.
    
\begin{table*}
\caption{\textrm{The table shows the expected CCSN rates within the SDSS DR5
    spectroscopic survey area (14\% of the entire sky) out to a distance of \mbox{$z\sim0.04$}. }}\label{SFG_rate_table}
  \begin{center}
    \begin{tabular}{c c c c c c c}
      \multicolumn{7}{c}{\bf HSFG Catalogue}\\
      \hline\hline
      & \multicolumn{6}{c}{\bf INDIVIDUAL GALAXY SN RATE LIMITS}\\
      & \multicolumn{2}{c}{\bf $>$ 0.0 SNe yr$^{-1}$} & \multicolumn{2}{c}{\bf $>$ 0.001 SNe yr$^{-1}$} & \multicolumn{2}{c}{\bf $>$ 0.01 SNe yr$^{-1}$}\\
      \hline
      {\bf 12+log(O/H)} & {\bf Galaxies} & {\bf SNe yr$^{-1}$} & {\bf Galaxies} & {\bf SNe yr$^{-1}$} & {\bf Galaxies} & {\bf SNe yr$^{-1}$}\\
      \hline
      {\bf No Limit} & 18\,350 & 115.6 & 13\,974 & 113.4 & 2\,557 & 73.6\\
      {\bf $<$ 8.4} & 8\,019 & 13.2 & 3\,650 & 11.2 & 120 & 2.3\\
      {\bf $<$ 8.3} & 4\,290 & 6.9 & 1\,830 & 5.9 & 73 & 1.3\\
      {\bf $<$ 8.2} & 1\,713 & 3.1 & {\bf 727} & {\bf 2.8} & 42 & 0.8\\
      {\bf $<$ 8.1} & 537 & 1.0 & 209 & 0.9 & 16 & 0.3\\
      \hline
      \multicolumn{7}{c}{}\\
      \multicolumn{7}{c}{\bf LSFG Catalogue}\\
      \hline\hline
      & \multicolumn{6}{c}{\bf INDIVIDUAL GALAXY SN RATE LIMITS}\\
      & \multicolumn{2}{c}{\bf $>$ 0.0 SNe yr$^{-1}$} & \multicolumn{2}{c}{\bf $>$ 0.001 SNe yr$^{-1}$} & \multicolumn{2}{c}{\bf $>$ 0.01 SNe yr$^{-1}$}\\
      \hline
      {\bf 12+log(O/H)} & {\bf Galaxies} & {\bf SNe yr$^{-1}$} & {\bf Galaxies} & {\bf SNe yr$^{-1}$} & {\bf Galaxies} & {\bf SNe yr$^{-1}$}\\
      \hline
      {\bf No Limit} & 6\,000 & 34.1 & 3\,691 & 33.2 & 901 & 22.9\\
      {\bf $<$ 8.4} & 1\,757 & 0.8 & 116 & 0.3 & 6 & 0.1\\
      {\bf $<$ 8.3} & 1\,025 & 0.3 & 50 & 0.1 & 0 & 0.0\\
      {\bf $<$ 8.2} & 401 & 0.1 & 18 & 0.0 & 0 & 0.0\\
      {\bf $<$ 8.1} & 116 & 0.0 & 4 & 0.0 & 0 & 0.0\\
      \hline 
      \multicolumn{7}{c}{}\\
      \multicolumn{7}{c}{\bf Nearby Bright Galaxy Catalogue}\\
      \hline\hline
       {\bf } & {\bf } & {\bf Galaxies} & {\bf } & {\bf SNe yr$^{-1}$} & {\bf } & {\bf }\\
      \hline
      {\bf } &  & 1 216 &  & 12.70 &  & \\
        \hline 

    \end{tabular}  
  \end{center}
\end{table*}
  
  For the purpose of this survey strategy we decide that the galaxy catalogue that
  optimally fits our requirements is the sub-sample of the HSFG cataglogue that
  contain galaxies with oxygen abundances of less than 8.2
  dex {\it and} a CCSR greater than 1 CCSNe every 1\,000 years. This low-metallicity catalogue contains
  727 galaxies, a suitable number for a pointed survey, with an estimated
  CCSR of \mbox{2.8 CCSNe yr$^{-1}$}. It should be noted that this galaxy catalogue is extracted solely from the 5\,713 square degrees of the sky that
  SDSS DR5 spectroscopic survey covers, that is 14\% of the entire sky.
  It can be assumed that the rest of the sky contains similar density of
  these low-metallicity galaxies. We shall return to this point later.
  
\subsection{Monte-Carlo Simulations}\label{sec_e}

  Of the estimated \mbox{2.8 CCSNe yr$^{-1}$} that this
  low-metallicity galaxy catalogue will produce, we will only be able to detect a fraction
  of these due to the practical limiting factors of a pointed survey. The
  reason that a given SN would not be detected is simply due to its faintness at the point of
  observation. The factors that will influence the
  likelihood that
  a SN will be detected when observed within our search are: whether
  or not the galaxy that hosts the
  CCSN is observable during the period of time when the CCSN
  is
  detectable, the type of CCSN (IIP, IIL, Ib, Ic or IIn) observed and its intrinsic
  brightness, the distance to the host galaxy, the extinction towards the
  CCSN, the exposure time and the age of the CCSN when observed (affected by the
  cadence of the observations).
  
  By running a Monte-Carlo simulation, constrained by each of these parameters,
  to randomly produce a sample of 100\,000 possible SNe observable
  within our search, we can infer the fraction of CCSNe that we should
  actually detect.
  
\subsection{Supernova Rates, Template Lightcurves and Distributions}

  In order for the Monte-Carlo simulation (MCS) to accurately reproduce the
  relative rates of the different types of CCSNe, we
  use the observed rates compiled by \citet{smartt_et_al}, given in Table
  \ref{smartt_rate}. These rates have been compiled within a time and volume-limited
  sample, accounting for all SNe discovered within the eight year period between 1999 January 1
  and 2006 December 2006 in galaxies with a recessional velocity \mbox{$< 2\,000
  {\rm km s}^{-1}$}. Apart from those SNe that
  inhabit environments of heavy extinction or first observed late into their
  evolution, it is expected that within this appointed distance limit 
  \mbox{($\mu = 32.3$)} all known types of SNe should have been bright enough to
  have been detected, implying that these relative rates are
  as free from any Malmquist bias as possible. CCSNe of type IIb have been merged with type Ib/c,
    and type IIn have been divided between those with a plateau phase in the
    tail of their lightcurves, IIn/P, and those with a linear phase, IIn/L.
  
\begin{table}
 \caption{{\textrm Relative CCSRs taken from
    Smartt et al. (2007).}\label{smartt_rate}}
  \begin{center}
    \begin{tabular}{ c c c c}
      \hline\hline
      {\bf SN Type} & {\bf Number} & {\bf Relative Rate} & {\bf Core-Collapse
      Only}\\
      \hline
      {\bf Ia} & 25 & 24.8$\%$ & -\\
      {\bf IIP} & 43 & 42.6$\%$ & 56.6$\%$\\
      {\bf IIL} & 2.5 & 2.4$\%$ & 3.3$\%$\\
      {\bf Ib/c} & 28 & 27.7$\%$ & 36.6$\%$\\
      {\bf IIn/P} & 1.25 & 1.2$\%$ & 1.6$\%$\\
      {\bf IIn/L} & 1.25 & 1.2$\%$ & 1.6$\%$\\
      \hline
    \end{tabular}
  \end{center}
\end{table}
  
\begin{figure*}
  \begin{center}
  \includegraphics[totalheight=0.3\textheight]{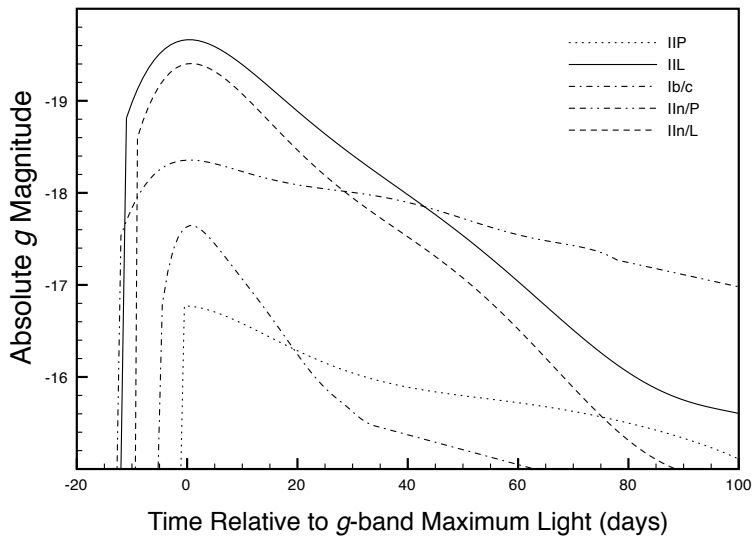}
  \caption{\textrm{Template lightcurves used within the Monte-Carlo
  simulation. Lightcurves for SNe of type IIP are according to 1999em, type IIL to 1998S, type Ib/c to 2002ap, type IIn/P to 1994Y and
  type IIn/L to 1999el. Note that all magnitudes are in the AB system.}}
  \label{templates}
  \end{center}
\end{figure*}

  We also supply template lightcurves of the various SNe for the MCS. For Type IIP we use
  1999em as our template, taking the data from \citet{2001ApJ...558..615H}. For Type IIL we use 1998S,
  taking data for the rise to maximum light from \citet{2000A&AS..144..219L} and
  data for the tail from \citet{2000MNRAS.318.1093F}. For Type Ib/c we use
  2002ap, taking data from \citet{2003PASP..115.1220F}. For
  SNe of Type IIn, we suggest that it is appropriate to divide the relative rate evenly
  between Type IIn that exhibit a plateau phase in their
  lightcurves and those that exhibit a linear phase, Type IIn/P and Type
  IIn/L respectively. For the Type IIn/P, we use
  1994Y as the template taking data from \citet{2001PASP..113.1349H}, allowing
  1998S to provide the rise to maximum light. For Type IIn/L we use 1999el as the
  template taking data from \citet{2002ApJ...573..144D}, again allowing 1998S
  to provide the rise to maximum light (see Fig \ref{templates} for comparison). 
  The following conversion from \citet{2007AJ....133..734B} is used to transform the data for the
  lightcurves to the Sloan $g$-band (note magnitudes are to be in AB system):

\begin{center}  
\begin{equation}
  g=B-0.03517-0.3411(B-V)
  \label{blanton_equ}
\end{equation}
\end{center}  

  The SN absolute magnitude distributions of \citet{2002AJ....123..745R} are used in the MCS to provide weighted,
  random distributions of peak magnitudes for the
  SNe (Table \ref{rich_table}). Equation \ref{blanton_equ} is again
  used to transform these distributions to the $g$-band, taking
  a $(B-V)$ colour from the epoch of peak $g$-magnitude from our template
  lightcurves.  $\sigma$ is the range in the peak magnitude distribution and $g$-band magnitudes are given in the AB system. When choosing a filter to perform a SN search with the Liverpool telescope the $r$-band is superior to the $g$-band because it has a greater filter throughput, the CCD detector is more responsive in the $r$-band and also SNe are generally brighter in the $r$-band especially later in their evolution. However, the \citeauthor{2002AJ....123..745R} distributions of SN peak magnitudes are given in the $B$-band which we can transform to the $g$-band but not to the $r$-band. It is for this reason that we simulate a SN survey in the $g$-band and then with these results we can then hypothesise the outcome of a search in the $r$-band.
  
\begin{table}
\caption{\textrm{Peak magnitude distributions from \citet{2002AJ....123..745R}}\label{rich_table}}
  \begin{center}
     \begin{tabular}{c c c c}
     \hline\hline
     {\bf SN Type} & ${\bf M_{B}}$ & ${\bf M_{g}}$ & ${\bf \sigma}$\\
     \hline
     {\bf IIP} & -17.00 & -17.02 & 1.12\\
     {\bf IIL} & -18.03 & -18.00 & 0.90\\
     {\bf Ib/c} & -18.04 &-18.22 & 1.39\\
     {\bf IIn/P} & -19.15 & -19.24 & 0.92\\
     {\bf IIn/L} & -19.15 & -19.24 & 0.92\\
     \hline
     \end{tabular}
   \end{center}
\end{table}

\subsection{Cadence, Distance and Extinction}

  A major factor influencing the detection efficiency of any SN
  search is the time between consecutive
  observations of a galaxy hosting a SN. As this
  time increases, so does the likelihood that we will not observe the
  SN until it is well advanced along its lightcurve. This could result
  in the magnitude dropping below our detection limit, hence failing to be detected in the
  search. There are three parameters that affect the cadence of our
  search and
  that need to be included in the MCS. The first of these is the probability
  that the SN will be in solar conjunction for a
  fraction, if not all, of the time that it
  remains detectable by our search. To account for this factor, we use
  the celestial coordinates of the 727 galaxies within our low-metallicity galaxy catalogue and an almanac for the Liverpool Telescope to determine the fraction
  of the year that each individual galaxy is observable, producing a
  distribution with a mean fraction of 0.380. We then use this
  distribution in the MCS to allocate a random amount of time to the
  cadence of observation of each SN, to account for cases where a SN
  would be unobservable and hence undetectable because its host galaxy is behind the sun.
  
  The second factor that affects the cadence is the
  number of nights that cannot be used by the Liverpool Telescope to
  perform our search. Reasons why the telescope could not be used on any
  particular night include weather, technical
  difficulties and scheduled maintenance nights. To
  compensate for these nights in our MCS, we have taken nightly reports
  from the Liverpool Telescope website\footnote{LT website:
  \it{http://telescope.livjm.ac.uk/}}, from 2005 August 1 to 2006 July 31,
  and use these reports to determine a distribution of nights (58$\%$ of nights in total) that
  can be used for our search. This distribution of usable nights is included in our MCS when considering the cadence of observations.
  
  The final factor that will affect the cadence of observations, is the
  number of galaxies that we aim to observe every night. This number is
  influenced by the amount of telescope time dedicated to our
  search each night; the
  greater the amount of time on the telescope each night, the greater the number of galaxies we can observe per
  night and the higher the cadence of observations. The number of galaxies
  that can be observed each night is also affected by exposure time. The
  benefit of increasing the exposure time is that we can search to a deeper magnitude limit, meaning
  that fainter SNe are detectable and SNe are detectable for a greater period of time. It is essential therefore to run a number of MCSs in order to find
  the optimal balance between the exposure time given to each galaxy and
  the cadence of observations and hence enabling us to detect the greatest fraction of
  CCSNe hosted by our low-metallicity galaxy catalogue.
  
  The final two parameters placed in our MCS are the host galaxy distances and
  the extinction toward the SNe. The MCS will attribute to each SN a random host galaxy
  distance, out to the distance limit of our search, which is \mbox{z$=$0.04} (assuming the galaxies within
  our search are spread homogeneously and isotropically throughout this
  volume). Considering that the majority, if not all, of the galaxies
  within our catalogue are low-luminosity, blue, dwarf galaxies with relatively
  low metallicities, we assume that host galaxies will have produced very
  little dust and will not contribute a great deal to any extinction toward
  the CCSN.
  To this end we simply attribute a typical Galactic extinction to each
  SN \mbox{(A$_{g}=0.3$)}.
  
  We note that a fraction of SNe remain undetected within spiral galaxies that are not observed face on \citep[e.g. ][]{2003astro.ph.10859C}. This is due to the fact that the average extinction attributed to SNe in spiral galaxies with higher inclination angles is higher than that of face on spiral galaxies, making the SNe fainter and more difficult to detect. Implementation of a correction for this effect into the MCS, using a similar method to that of \citet{2005MNRAS.362..671R}, could easily result in an under-estimation of the CCSRs observed in low-metallicity galaxies as these tend not be to grand-design spiral galaxies but rather dwarf, irregular galaxies where the effect of galaxy inclination would be negligible. We have therefore decided not to attempt to correct for this effect. This may make our
predictions for the SN discovery rates in the full galaxy catalogues
(with no metallicity limit) somewhat optimistic. However this may not lead to a significant over-estimate of events, as the SFRs that we calculate come from the observed galaxy H$\alpha$ luminosities and in inclined spirals these will also be lower than in face on targets.

\subsection{Monte Carlo Simulation Results for a Single 2.0m Telescope}\label{sec_f}

  Having included each of these parameters in the MCS, we randomly
  generate 100\,000 SNe that would potentially be observable within
  our search. The information gained about each SN produced by the simulation include
  its type and its apparent magnitude at the point of observation. It is only if this magnitude is
  above the limiting magnitude of our search can we register the SN as being
  detected.
  
\begin{figure*}
  \begin{center}
  \includegraphics[totalheight=0.3\textheight]{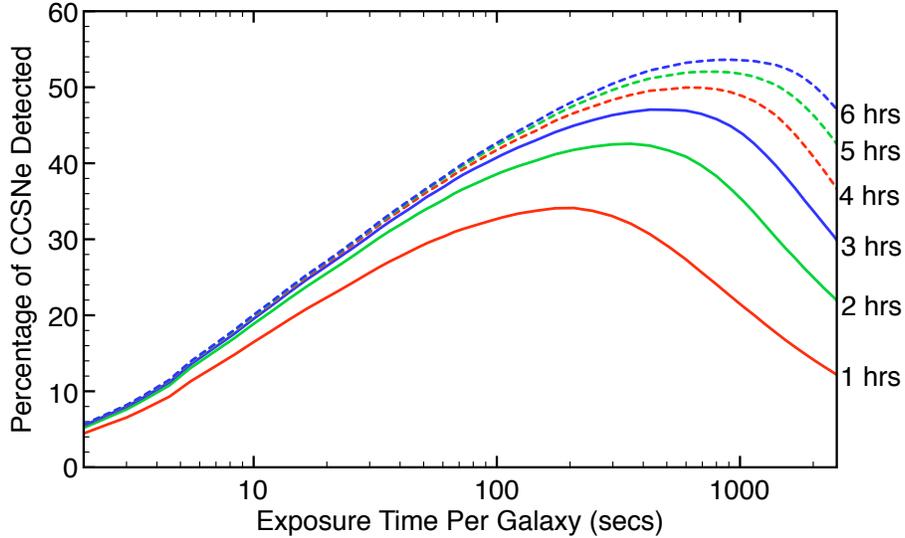}
  \caption{\textrm{Percentage of CCSNe
  detected from our MCSs for a differing number of hours observing per night
  (labeled to the right of the plot). The percentage of CCSNe detected
  increases as the nightly observing time increases, as this increases the
  cadence of observations. As the individual exposure time per galaxies
  increases, the limiting magnitude of the search deepens but the cadence of
  observations decreases. There is therefore an optimal point at which
  the exposure time and cadence of observation are balanced to detect the maximum
  percentage of SNe.}}
  \label{exp_time}
  \end{center}
\end{figure*}

  By running the MCS multiple times for (a) the varying number of hours we aim to
  observe with the LT every night and (b) the varying amount of exposure time
  that we dedicate to each individual galaxy, we can determine the optimal
  values of these parameters that shall enable us to detect the greatest
  fraction of the CCSNe that we have predicted our low-metallicity galaxy
  will produce. The results can be seen in Fig. \ref{exp_time}.
  
  Studying the results of the MCSs we suggest that the
  difference between the percentage of CCSNe detected while aiming to observe one
  hour per night as opposed to two hours per night or more is not enough to
  justify the extra observing time as the
  fraction of CCSN detected only increases by a few percent. As the detection rates are double valued either side of the peak
  detection rate, we suggest using a shorter exposure time in order to
  increase cadence and thereby increase the probability of 
  discovering CCSNe earlier in their evolution as opposed to later. 
  
   Aiming to observe one hour a night using 60 sec exposures allows each galaxy to be observed once every $\sim$17 days while accessible with LT and enables 29.6\% of CCSNe to be detected with
  our search. When considering the catalogue of 727 low-metallicity galaxies which produce \mbox{2.8 CCSNe yr$^{-1}$}, this equates to a detected CCSN rate of \mbox{0.8 SNe yr$^{-1}$}
  and requires a total amount of \mbox{154hrs
  yr$^{-1}$} telescope time. Remembering that the $r$-band is superior to the $g$-band for performing a SN search, we note that given a 60 sec exposure the AB limiting magnitudes for the Liverpool telescope are $g = 18.7$ mags and $r = 19.0$ mags. Given the average cadence of observations determined by our chosen survey parameters, a typical SN will be observed $\sim$73 days post-explosion with $g-r=0.8$ mags. Given the difference in limiting magnitudes and the typical $g-r$ colour at the point of observation, we expect to detect SNe that are 1.1 magnitudes fainter with an $r$-band search than we would with a $g$-band search. Assuming that the template SN lightcurves and the distribution of SN peak magnitudes are similar in the $r$-band as compared to the $g$-band, we re-run the $g$-band MCS with a limiting magnitude 1.1 magnitudes fainter than previously used. From this simulation we predict that using the $r$-band will allow for $47.8\%$ of CCSNe to be detected with our search. Again considering the catalogue of 727 low metallicity HSFGs which produce \mbox{2.8 CCSNe yr$^{-1}$}, this equates to a detected CCSN rate of \mbox{1.3 SNe yr$^{-1}$}.
  
  Assuming a typical Galactic extinction of A$_g=0.3$ (A$_r=0.22$) and an exposure
  time of 60 seconds we estimate that the absolute
  limiting magnitude ($25\sigma$ significance) for this search at a distance of 20Mpc will be
  $M_r=-12.7$, at 70Mpc it will be $M_r=-15.4$ and at 150Mpc it will be
  $M_r=-17.1$. We plan to use the method of image matching and subtraction \citep[see ][]{1998ApJ...503..325A} to detect SNe throughout this survey. Assuming original 
images with equal depth, the process of image subtraction increases the
noise by roughly a factor of $\sqrt{2}$. In addition, the detection of
SNe inside their host galaxies will be always more difficult due to
image subtraction residuals caused by uncertainties in the alignment and
matching of the two images. Since we wish to avoid a large fraction of 
spurious detections, we require the relatively high significance level of 
25$\sigma$ from the limiting magnitude of our search.

\subsection{Surveying with a Network of 2.0m Telescopes}\label{sec_f2}

  As discussed previously SDSS DR5 covers only $\sim$14\% of the
  entire sky. If we presently had the ability to compile a catalogue of low-metallicity
  galaxies for the whole sky as complete as that for the area of the SDSS DR5
  spectroscopic survey and a collection of seven 2.0m telescopes (or
  three to four 2.0m telescopes with double the time allocation) we would expect to detect
  roughly 7 times the CCSNe detected solely by the LT, i.e. $\sim$9.3
  CCSNe/yr. We could consider using a network of six to eight 2.0m robotic telescopes similar to the RoboNet global network of 2.0m
  telescopes consisting of the LT and the Faulkes Telescope North and the
  Faulkes Telescope South to perform this kind of survey. Another advantage of using a network of telescopes in both the northern and southern hemispheres as opposed to a single telescope is that we gain a greater sky coverage and can therefore target a greater number of galaxies within our search.
  
  The two obstacles that we would encounter if we were to use this strategy to search for CCSNe are firstly the generous amount of telescope
  time that we would require \mbox{($\sim$1\,000 hrs yr$^{-1}$)} and secondly the present lack of
  data required to compile an all sky galaxy catalogue. Compiling a catalogue
  of all galaxies listed in the 2dF, 6dF, LEDA and SDSS DR5 within the redshift range
  \mbox{0$<$z$<$0.04} results in a catalogue containing a total of 103\,549 individual galaxies, only a fraction of which will be star-forming. Apart from this we shall have to wait for PS1, the prototype
  1.8m telescope of the Panoramic Survey Telescope and
  Rapid Response System (Pan-STARRS), which shall cover an area of 3$\pi$ of
  the sky to a depth exceeding that of SDSS DR5, in order to compile a far more
  complete all-sky low-metallicity galaxy catalogue.

\section{Strategy 2 : volume-limited searches with the Pan-STARRS all sky surveys} \label{sec_g}

  Having considered a dedicated pointed low-metallicity galaxy survey to search for CCSNe both with one and with a network of 2.0m telescopes, we
  now turn our attention to an all-sky survey. Pan-STARRS is a
  set of four 1.8m telescopes with wide field optics, each with a 7 square
  degree field of view, which will be constructed by the University of Hawaii. The
  prototype telescope PS1 has now achieved first light and
  is set to go online during 2008. PS1 will have the capability of
  surveying
  the whole available sky in less than a week at a rate of 6\,000 square degrees per
  night, covering 30\,000 square degrees of the sky per cycle
  \citep{PS1_DRM}.
    Included in the list of tasks that the PS1 image reduction pipeline will
  perform is the subtraction of every image from a reference image in order to
  produce a database of residual transient objects that will include moving objects and static
  variables such as SNe.

There are two different strategies that one would employ to search for
SNe (or transients of any type), these are volume-limited searches and
magnitude-limited searches. We will consider both of these. A volume-limited search has the advantage that it can quantify true rates of 
transients. In addition the radial limit can be chosen so that the 
target discoveries are bright enough to be followed in multi-wavelength
studies with complementary facilities (e.g. spectroscopic and photometric
follow-up with 2-8m telescopes).

\subsection{Monte Carlo Simulations}\label{sec_h}

   Of all the modes in which PS1 shall be run the 3$\pi$ Steradian Survey
   (covering 30,000 square degrees of the sky)
   shall be the most effective when searching for nearby CCSNe. The survey aims to cover that whole sky 60 times in 3 years, that is 12 times
   in each of the five filters {\it g, r, i, z} and {\it y}. This aim has already
   taken historic weather patterns on Haleakala into consideration. The footprint of the survey
   completely covers the entire footprint of the SDSS DR5 spectroscopic
   survey and the AB limiting magnitude \mbox{(25$\sigma$)} in the $g$-band is stated
   to be 21.5
   mags for a single 60 second exposure \citep{PS1_DRM}.
   
   Having an adapted MCS produce 100\,000 potential SNe over the
   redshift range of our galaxy samples \mbox{($0<z<0.04$)} will again help us
   to estimate the fraction of potentially observable CCSNe that we will
   actually detect with PS1. We assume that we can take the
   fraction of CCSNe detected in the $g$-band as the typical fraction that
   would be detected in each of the five filters; indeed running the adapted MCS detailed in Section \ref{sec_i} in an attempt to produce the detected CCSN rates of this nearby sample in all five filters, were the {\it K}-corrections essentially behave as colour corrections at such low redshift, results in predicted detection efficiencies extremely close to those found if considering only the $g$-band. Concerning the cadence we have also taken the
   PS1 survey strategy into consideration \citep{PS1_DRM}, see
   Figure \ref{cad_fig}. 
A random extinction is assigned to each of the CCSNe
   using a weighted distribution based on CCSRs taken from \citet{2001MNRAS.324..325M} and
   \citet{1994ApJ...432...42S}. However for 
   low metallicity galaxies ($\leq$8.2\,dex) 
  only a typical Galactic extinction of \mbox{A$_{g}=0.3$}
   is assigned to the CCSNe, as at these metallicities we expect the galaxies
   to be relatively dust free.
   
\begin{figure}
  \begin{center}
  \includegraphics[totalheight=0.25\textheight]{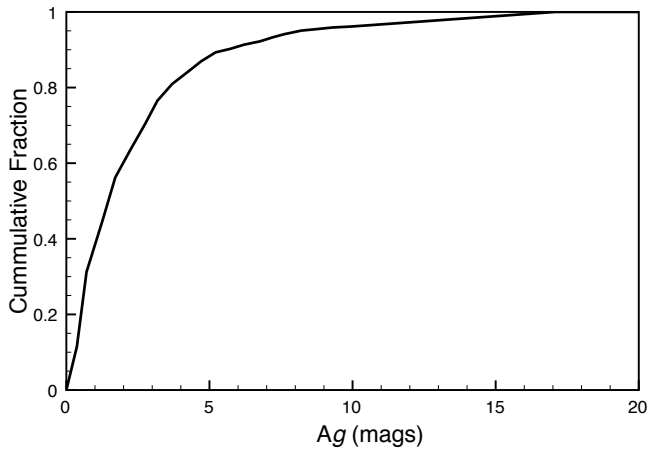}
  \caption{\textrm{Cumulative distribution of ~1\,500 extinctions as measured through the fraction of the Galactic Plane which lies within the Pan-STARRS survey area.}}
  \label{ext_dist}
  \end{center}
\end{figure}

  We must also consider the fact that Pan-STARRS shall be covering a large fraction of the sky that is obscured by the Galactic Plane, a total of $\sim$7\,500 square degrees of the Pan-STARRS area. Any SN that is located with a line-of-sight through the Galactic Plane shall be extinguished above the typical Galactic extinction assumed. To allow for this we take the extinctions, as determined by \citet{1998ApJ...500..525S}, of $\sim$1\,500 points positioned homogeneously across the expanse of the Galactic Plane visible with Pan-STARRS and determine a distribution of extinctions to be folded into the MCS for any SN falling behind the Galactic Plane. The cumulative distribution of these extinctions can be seen in Figure \ref{ext_dist}.
   
   For the full galaxy catalogue (within $z \leq 0.04$) 
  we can expect to detect \mbox{66.9$\%$} of all
   CCSNe. When considering the lower metallicity samples we can expect to detect
   74.2$\%$ of CCSNe. The predicted detectable CCSRs for both the full galaxy catalogue
   and for the lower-metallicity samples can be found in Table \ref{PS1_rates}.
   
   These results are a vast improvement on the rates predicted for the
   catalogued galaxy surveys with both a single 2.0m telescope and a network
   of 2.0m telescopes. Considering that the Pan-STARRS survey area is
   $\sim$5.25 that of SDSS DR5 and given a complete galaxy catalogue of that
   area, we can expect to detect a total of \mbox{$\sim$570 CCSNe yr$^{-1}$}, roughly 13 of which
   will be from host galaxies with \mbox{$12+{\rm \log(O/H)}<8.2$}. 
We would
expect that the PS1 survey itself will provide a photometric catalogue
of at least the quality the depth of SDSS DR5 (most likely signficcantly
better) over the whole sky and hence provide us with a means to identify
SNe in nearby, faint blue hosts. 
Our estimate of 
   number of CCSNe expected to be detected in the low-metallicity galaxy sample can be considered as a lower
   limit as there may be many more blue, compact, dwarf galaxies that have
   not been detected and catalogued by SDSS DR5.
 The number of CCSNe expected to be detected
   by PS1 in low-metallicity galaxies is hence about 50\%
 times more than would be possible using a network of seven 
2.0m telescope in a dedicated pointed survey with 1000\,hrs of dedicated
time. It would this seem much more efficient to use the PS1 survey
to discover these are objects than invest in a dedicated pointed survey
with large numbers of 2.0m telescopes with standard CCD cameras
and fields of view. 
   
\subsection{PS4}

   Eventually three more 1.8m telescopes identical to PS1 will be added to
   the Pan-STARRS network to create PS4. These four telescopes will be trained
   on the same area of the sky simultaneously 
   to effectively perform a deeper all-sky
   survey (with the same observing strategy as PS1), 
giving a co-added 25$\sigma$ limiting magnitude of $g$=22.2.
   
   At an \textit{apparent} limiting magnitude of \mbox{$m_{g}$=21.5}, PS1 has an \textit{absolute} limiting
   magnitude \mbox{$M_{\rm g}$=-14.7} at the boundary of our search ($z=0.04$). For
   PS4 we would push the boundary of our search out to the point where the
   absolute limiting magnitude is identical to that of our search with PS1 in
   order to detect the same fraction of CCSNe as we have predicted for PS1.
   This distance limit is calculated to be $z=0.056$.
   Using the same 
   survey strategy as previously suggested for PS1 and introducing
   the new limiting magnitude and distance limit, we predict that PS4 should discover a total of \mbox{$\sim$1\,308 CCSNe yr$^{-1}$}, \mbox{$\sim$18.3 CCSNe yr$^{-1}$} being found in galaxies with \mbox{$12+{\rm \log(O/H)}<8.2$}. These results
   can be found in Table \ref{PS4_rates}.
   
   \begin{table}
\caption{\textrm{Estimated intrinsic and detectable CCSRs within the SDSS
    DR5 survey area and $z<0.04$. Also shown are the rates extended to the PS1 survey area.}\label{PS1_rates}}
  \begin{center}
    \begin{tabular}{c c c c c}
      \hline\hline
      & \multicolumn{2}{c}{\bf Total Sample} & \multicolumn{2}{c}{\bf Detected CCSNe yr$^{-1}$}\\
      \hline
      {\bf 12+log(O/H)} & {\bf Galaxies} & {\bf CCSNe yr$^{-1}$} & {\bf SDSS Area} & {\bf PS1 Area}\\
      \hline
      {\bf No Limit} & 25\,091 & 162.4 & 108.6 & 570.4\\
      {\bf $<$ 8.4} & 9\,776 & 13.9 & 10.3 & 54.1\\
      {\bf $<$ 8.3} & 5\,315 & 7.2 & 5.3 & 28.0\\
      {\bf $<$ 8.2} & 2\,114 & 3.2 & 2.4 & 12.5\\
      {\bf $<$ 8.1} & 653 & 1.1 & 0.8 & 4.3\\
      \hline
    \end{tabular}  
  \end{center}
\end{table}

\begin{table}
\caption{\textrm{Estimated intrinsic and detectable CCSRs within the SDSS
    DR5 survey area and $z<0.056$. Also shown are the rates extended to the PS4 survey area.}\label{PS4_rates}}
  \begin{center}
    \begin{tabular}{c c c c c}
      \hline\hline
      & \multicolumn{2}{c}{\bf Total Sample} & \multicolumn{2}{c}{\bf Detected CCSNe yr$^{-1}$}\\
      \hline
      {\bf 12+log(O/H)} & {\bf Galaxies} & {\bf CCSNe yr$^{-1}$} & {\bf SDSS Area} & {\bf PS4 Area}\\
      \hline
      {\bf No Limit} & 42\,335 & 372.4 & 249.1 & 1\,308.0\\
      {\bf $<$ 8.4} & 12\,104 & 24.6 & 18.3 & 95.8\\
      {\bf $<$ 8.3} & 6\,053 & 11.6 & 8.6 & 45.2\\
      {\bf $<$ 8.2} & 2\,309 & 4.7 & 3.5 & 18.3\\
      {\bf $<$ 8.1} & 680 & 1.2 & 0.9 & 4.8\\
      \hline
    \end{tabular}  
  \end{center}
\end{table}

 \section{\textbf{Strategy 3 : magnitude-limited searches
with future all-sky surveys}}\label{sec_ii}
 
   Thus far when considering the Pan-STARRS survey to search for CCSNe
   we have limited ourselves to the known galaxy
   population within $z \lesssim 0.04$. If our aim is to detect
   \textit{all} of the CCSNe that we possibly can then we are by no
   means utilizing the Pan-STARRS survey to its greatest
   potential. The galaxy catalogue employed is
   taken from SDSS DR5 and is therefore limited by a comparatively
   shallow depth compared to the depth to which Pan-STARRS will be
   able to detect CCSNe \citep{PS1_DRM}.
   
   A far more powerful method that we could implement with
   Pan-STARRS to detect CCSNe is to perform simply a magnitude-limited search
   to detect CCSNe
   over the whole sky and to as large distances as possible
   following the example of surveys such as the
   {\it Nearby SN Factory} \citep{2006NewAR..50..436C}. The inherent
   difficulties that we will encounter with this kind of survey will
   be trying to distinguish SNe from all other transient events and
   variables such as classical novae, asteroids and AGN and further
   distinguishing CCSNe from the thermonuclear (Type Ia) SNe.   
   Assuming that it may be possible to overcome these
    difficulties (and that we can probabilistically classify transient events)
    we can
   estimate how many CCSNe that we shall be able to detect with PS1, PS4 and
   also compare these rates with the likes of the Large Synoptic Survey
   Telescope (LSST) set to come online in 2016.
   
\subsection{Adapted Monte-Carlo Simulations}\label{sec_i}

\begin{figure}
  \begin{center}
  \includegraphics[totalheight=0.2\textheight]{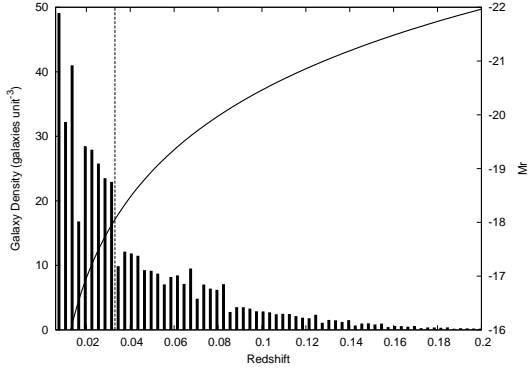}
  \caption{\textrm{The galaxy distribution for the SDSS DR5 spectroscopic
  survey galaxy sample. The solid curve represents the absolute magnitude
  limit as a function of redshift given that the SDSS DR5 apparent magnitude
  limit for the sample is $r=17.77$. An absolute magnitude limit of \mbox{$M_{\rm
  r}=-18$} is reached at at redshift of 0.033 (dashed line). There is a clear
  drop in the galaxy density at this redshift, indicated by the dashed line,
  and we choose to use this limit as the completeness limit of the SDSS DR5
  galaxy sample.}}
  \label{sdss_galaxies}
  \end{center}
\end{figure}

   In order to determine the total CCSRs that we are to expect from the all-sky
   surveys PS1, PS4 and LSST, we must first ascertain to what redshift depth our SDSS
   DR5 galaxy sample is relatively complete to. When targeting galaxies for the
   spectroscopic survey SDSS DR5 has a lower magnitude limit of $r=17.77$.
   This apparent magnitude limit equates to an absolute galaxy magnitude
   limit of \mbox{$M_{\rm r}=-18$} being reached at a redshift of $z=0.033$. The
   galaxy density distribution for SDSS DR5 can be seen in Figure
   \ref{sdss_galaxies}. There is a clear drop in the observed galaxy density at this
   redshift and we choose to set this limit as the completeness limit of the
   SDSS DR5 galaxy sample.
   
\begin{figure}
\begin{center}
\includegraphics[totalheight=0.3\textheight]{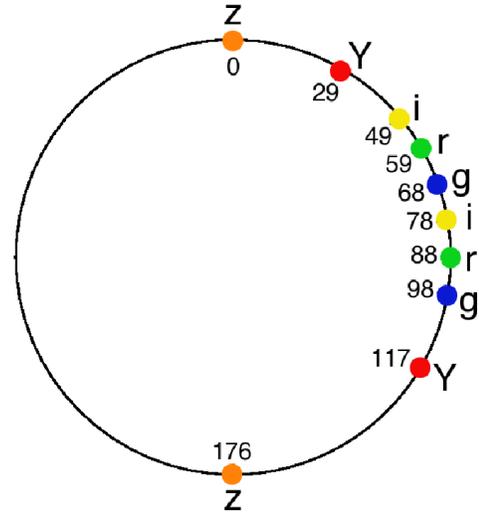}
\caption{\textrm{A schematic depicting the cadence of observation for any
given point in the sky, taken from the Pan-STARRS observation strategy
\citep{PS1_DRM}. One Pan-STARRS year of 12 lunar months, approximately 352
days, is represented by a circle. With each observation epoch a double exposure is taken in order to
identify Near Earth Objects}\label{cad_fig}}
\end{center}
\end{figure}

   With this completeness limit in mind we can now extrapolate the complete local
   galaxy sample out to the observable redshift limits for the individual CCSN
   surveys that we are considering. We assume that the galaxy
   distribution throughout the observable universe is both homogeneous and
   isotropic. To determine the observable redshift limits of each of the suggested
   CCSN searches it is necessary  to model the CCSN detection distribution
   for each of the surveys. This is made possible by using an adapted MCS to determine
   the percentage of CCSNe that would be detected as a function of redshift.
   The variant parameters included in the MCS that define one distribution from another are the
   cadence and the limiting magnitudes of the survey. For PS1 and PS4 we
   set the cadence to reflect the observing strategy for Pan-STARRS (Figure
   \ref{cad_fig}) and the 25$\sigma$ limiting magnitudes outlined by
   \citet{PS1_DRM}, assuming that the limiting magnitude of PS4 is
   achieved via co-adding four PS1 images (see Table \ref{mag_table}).
   
\begin{table}
\caption{\textrm{The 25$\sigma$ AB limiting magnitudes for PS1, PS4 and LSST}\label{mag_table}}
  \begin{center}
     \begin{tabular}{c c c c}
     \hline\hline
     {\bf FILTER} & {\bf PS1} & {\bf PS4} & {\bf LSST}\\
     \hline
     {\it g} & 21.5 & 22.2 & 22.7\\
     {\it r} & 20.9 & 21.7 & 22.4\\
     {\it i} & 20.8 & 21.6 & 21.8\\
     {\it z} & 19.8 & 20.6 & 21.2\\
     {\it Y} & 18.4 & 19.1 & 19.9\\
     \hline
     \end{tabular}
   \end{center}   
\end{table}

   The current LSST configuration sets the telescope at an 8.4m diameter with a 9.6
   square degree field-of-view. It is planned that LSST will survey the whole
   of the observable sky every 3 nights, eventually covering a total of 20\,000 square
   degrees. The primary focus of the survey is to detect transients and in
   order to detect fast faint events the sky will be surveyed with
   duplicate pairs of 15sec exposures. Using the LSST exposure calculator
   this equates to the 25$\sigma$ limiting magnitudes per single exposure as found in Table
   \ref{mag_table}.
   
   When considering a magnitude-limited survey it is important to note that we shall
   detect CCSNe at relatively large redshifts compared to the volume-limited survey that we have previously discussed. With this increase in
   survey depth we now encounter parameters in our search that we have
   previously neglected due to fact that our searches have been relatively
   nearby.These parameters that now have to be considered in our MCS are all
   functions of redshift and hence could be neglected at low-redshift. They are the
   star-formation history (SFH) of the universe, the redshifting
   of light emitted by the CCSNe, the broadening of the CCSNe
   lightcurves due to time-dilation and finally the retardation
   of the CCSN rates again due to time-dilation.

   {\it Star-Formation History:	} As star-formation was more prolific earlier
   in the Universe's history than it is at present, star-formation is seen to
   increase as redshift increases (at least to moderate redshifts of
   $z\sim2.5$). It is also important to note that approximately 50$\%$ of all local star formation activity is obscured, while at a redshift of z$\sim$1 this fraction could be as high as 80$\%$ \citep{2005A&A...440L..17T}. To incorporate the SFH of the universe into our simulations, but also taking into account that a greater fraction of this star-formation will be obscured by dust at higher redshifts, we include the far-ultraviolet (FUV) SFR evolution determined by \citet{2005ApJ...619L..47S}, see Equation (\ref{sfh_equ}). Compared with all other indicators of SFR, the FUV luminosity is the most inhibited by dust and therefore we would hope that by using this indicator we would only consider the star-formation that will give rise to CCSNe that are not so extinguished as to be invisible to our search.

\begin{equation}
  \dot{\rho}_{*}(z)=(1+z)^{2.5\pm0.7}.
  \label{sfh_equ}
\end{equation}

   {\it K Corrections:	} Comparing the bolometric fluxes of objects at various redshifts presents no
   difficulty as we are comparing the total flux emitted by the objects at all
   wavelengths. However, in practice we measure only a fraction of the total
   flux of an object, redshifted to the observed wavelength and transmitted through a given
   bandpass. This presents us with a challenge when we wish to simulate the
   total number of CCSNe that an all-sky survey will detect, as we have so
   far only considered their rest-frame (or emitted-frame) magnitudes but we
   must now consider their redshifted-frame (or observed-frame)
   magnitudes. To do
   this we need to calculate a transformation known as the $K$ correction
   \citep{1968ApJ...154...21O,2002astro.ph.10394H}, which transform between
   emitted-frame and observed-frame magnitudes:
   
\begin{equation}
m_R=M_Q+{\rm DM}+K_{QR}
\label{kcor_four}
\end{equation}
  
 \begin{figure*}[!b]   
\begin{equation}
  K_{QR}=-2.5{\rm log}_{10}\left[\frac{1}{[1+z]}\frac{\displaystyle{\int}\lambda_{obs}F_{\lambda}(\lambda_{obs})R(\lambda_{obs}){\rm d}\lambda_{obs}{\displaystyle{\int}\lambda_{em}S^Q_{\lambda}(\lambda_{em})Q(\lambda_{em}){\rm d}\lambda_{em}}}{\displaystyle{\int}\lambda_{obs}S^R_{\lambda}(\lambda_{obs})R(\lambda_{obs}){\rm d}\lambda_{obs}\displaystyle{\int}\lambda_{em}F_{\lambda}([1+z]\lambda_{em})Q(\lambda_{em}){\rm d}\lambda_{em}}\right]
 \label{kcor_nine}
\end{equation}
\end{figure*}

\noindent where $m_R$ is the apparent magnitude the the object as measured through observed-frame
bandpass, $M_Q$ is the absolute magnitude of the object as measured through the emitted-frame
bandpass, DM is the distance modulus of the object considered as
determined from its luminosity
distance and $K_{QR}$ is the $K$ correction that transforms between these
magnitudes. The $K$ correction can be understood as the difference in
magnitude between two identical objects placed at the same distance; one traveling at a redshift of
$z$ relative to the observer and measured through the observed-frame
bandpass, the other at rest relative to the observer and measured through the
emitted-frame bandpass.

We choose to use the $K$ correction as derived by \citet{2002astro.ph.10394H} to account for the redshifting of the SN light in our MCS; see Equation \ref{kcor_nine}, where \mbox{$F_{\lambda}(\lambda)$} is the spectral density of flux of the
object considered taken from spectral data (measured in \mbox{ergs cm$^{-2}$
s$^{-1}$ \AA$^{-1}$)}.
$R(\lambda)$ and $Q(\lambda)$ are the observed-frame and emitted-frame filter
total throughput functions, both
depending on the various parameters of the survey considered including the total
throughput of the atmosphere, the reflectivity of the telescope
mirrors, the transmission of the corrector optics and filters and the quantum
efficiency of the camera detector. $S^R_{\lambda}(\lambda)$ and
$S^Q_{\lambda}(\lambda)$ are the spectral density of flux for the AB zero-magnitude standard source, which
is a synthetic source equal to 3631 Jy (where \mbox{1 Jy = 10$^{-26}$ W m$^{-2}$
Hz$^{-1}$)} for all frequencies \citep{1983ApJ...266..713O}, as measured
through the observed-frame and emitted-frame filters respectively.

It is essential to provide our MCS with
$K$ corrections for each of the CCSNe simulated in order to determine
accurate apparent magnitudes as measured through each of the five Pan-STARRS
filters. As the $K$ correction varies for each filter considered we must
produce five $K$ corrections for each CCSN; $K_{gg},K_{gr},K_{gi},K_{gz}$ and $K_{gy}$. Any
given CCSN produced by the simulation is placed at a random redshift and is
first observed at a random point in its evolution, therefore we require $K$ corrections
over the entire redshift range of our surveys, for every possible epoch of
observation and for every CCSN type. It is then imperative that we have enough spectral information for each type of
CCSN considered in order to cover the shear expanse of the parameters
involved in calculating the $K$ corrections. We require spectral information
covering a large range of the evolution of each CCSN type including the rise
to maximum light and through to over 300 days post-explosion in order to
accommodate for the maximum time separating two consecutive observations in the same
filter according to the Pan-STARRS observation strategy implemented in the
MCS (see Figure \ref{cad_fig}). As the $K$ correction evolves with time we also require spectral data for a large
number of epochs in order to achieve good time resolution over the evolution
of each CCSN. Finally, a wide range in wavelength coverage is required in
order to provide enough information to determine $K$ corrections for all five
bandpasses whenever possible, remembering that the spectral data shall be
redshifted according to the distance appointed to the CCSN by the MCS.

For CCSNe of Type IIP we have used the spectra of SN1999em
\citep{2001ApJ...558..615H, 2001ApJ...553..861L, 2000ApJ...545..444B}, for
Type IIL the spectra of SN1998S are used \citep{2000MNRAS.318.1093F,
2004MNRAS.352..457P, 2005ApJ...622..991F} and for Type Ib/c the spectra of
SN2002ap are used \citep{2002MNRAS.332L..73G} supplemented with spectra from
SN1990B \citep[and Asiago Catalogue]{2001AJ....121.1648M} and SN1998bw
\citep{2001ApJ...555..900P} in order to attain the time coverage required. Due
to the fact that Type IIn/P and IIn/L are relatively rare, there is not enough
spectral data to cover both the time and wavelength coverage required to
determine a full range of $K$ corrections for these CCSNe. To compensate for
this lack of data we allow the Type IIn/P SN population to be represented
by Type IIP SNe and the Type IIn/L population to be represented by Type IIL
SNe, adapting the Smartt et al. (2007) rates appropriately for the MCS.

\begin{figure}
  \begin{center}
  \includegraphics[totalheight=0.25\textheight]{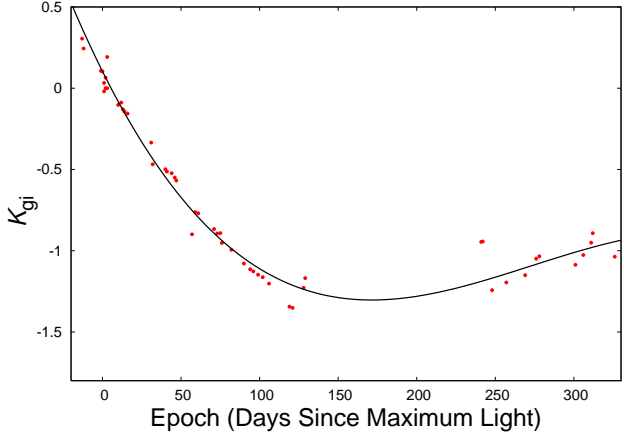}
  \caption{\textrm{A third order polynomial is fitted to each set of $K$
  correction data, $K_{gg},K_{gr},K_{gi},K_{gz}$ and $K_{gy}$, for each CCSNe
  type and each redshift position sampled over the entire redshift range of
  the all-sky surveys considered. This example is of the $K_{gi}$ correction
  for Type IIL SNe as sampled at a redshift of $z=0.2$}\label{kgi_fig}.}
  \end{center}
\end{figure}

Having calculated the thousands of possible $K$ corrections allowed by the
relevant parameters of the MCS, we then fit a third
order polynomial to each set of $K$ correction data defined by CCSN type,
filter and redshift in order to provide a $K$ correction for all epochs
required by the MCS (See Figure \ref{kgi_fig})

These polynomials are read by the MCS in order to determine a final accurate
apparent magnitude in each of the five filters, for each CCSN as measured at its specific redshift and
point in its evolution.

   {\it Time Dilation:	} The broadening of the CCSN lightcurves due to time dilation at relatively
   high redshift can be described simply as watching the CCSN
   evolve in `slow motion' with respect to our frame of reference. This
   enables us to sample the lightcurve of the CCSN with a greater frequency
   within its frame of reference relative to our frame of reference. The
   cadence hence goes as:
   
\begin{center}  
\begin{equation}
  \rm{cadence}=\rm{cadence}_{rest}\times(1+z).
\end{equation}
\end{center}
     
   The same argument leads to a retardation of the detected CCSN rates,
   giving:
   
\begin{center}  
\begin{equation}
  \rm{CCSR}=\rm{CCSR}_{rest}/(1+z).
\end{equation}
\end{center}

\begin{figure}
  \begin{center}
  \includegraphics[totalheight=0.25\textheight]{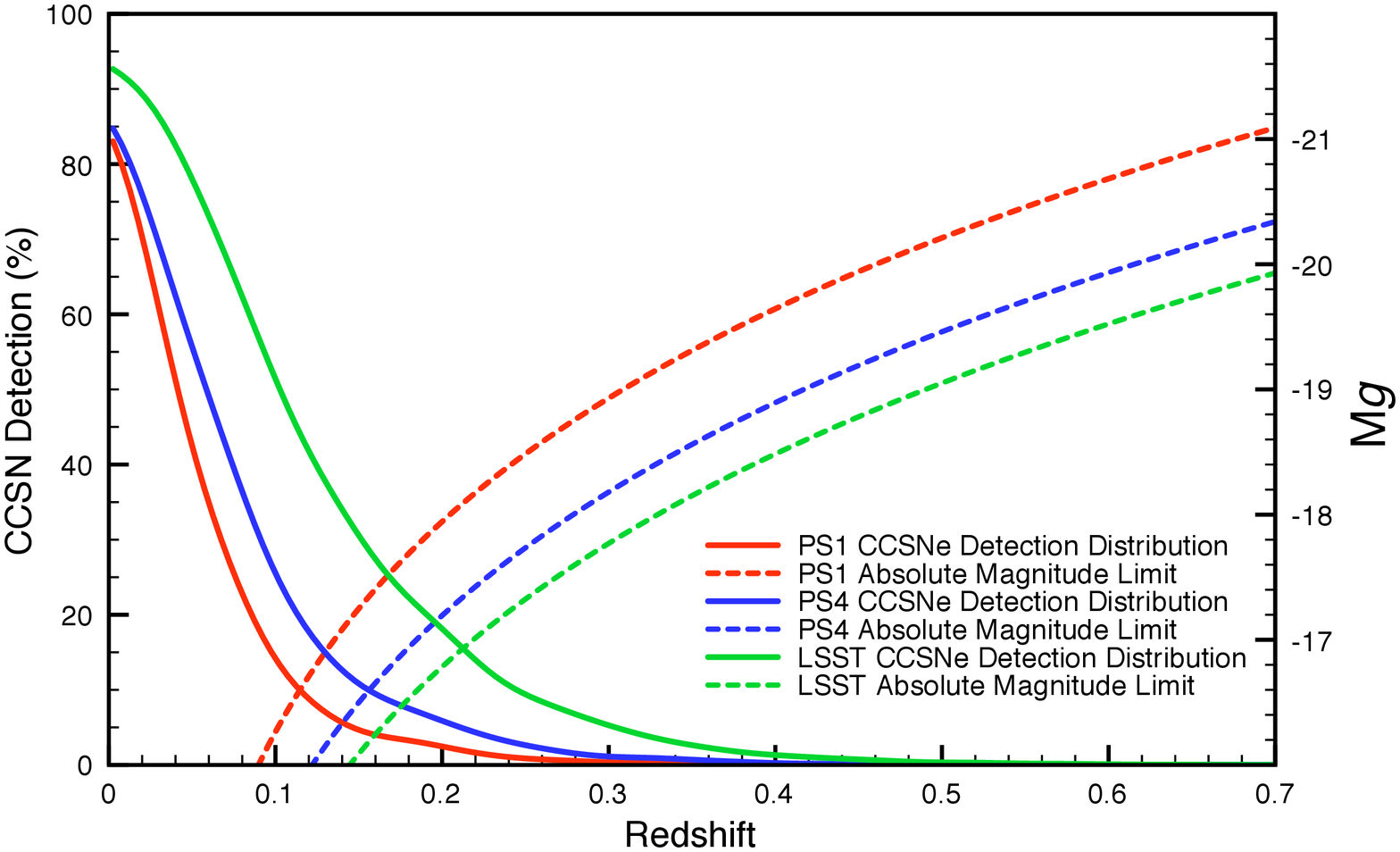}
  \caption{\textrm{The solid lines show CCSN detection distributions for PS1,
  PS4 and LSST, the dashed lines the absolute magnitude limits of the surveys
  as a function of redshift. At a absolute limiting magnitude of $M_{\rm
  g}=-20$ the PS1, PS4 and LSST reach their respective limiting redshifts
  depths of $z\sim0.4$, $z\sim0.5$ and $z\sim0.6$.}}
  \label{CCSN_distributions}
  \end{center}
\end{figure}

   Figure \ref{CCSN_distributions} shows the CCSN detection distributions for PS1, PS2
   and LSST. Beyond the redshift limit where an absolute limiting magnitude
   of $Mg=-20$ is attained the surveys will cease to detect all but the most
   extreme of CCSNe were luminosity is concerned and hence it is at this
   limit that we choose to mark the boundary of the observable universe when
   considering CCSNe. These redshift limits are 0.37, 0.50 and 0.59 for PS1,
   PS4 and LSST respectively.

   Assuming that the galaxy density is homogeneous and isotropic
   throughout the volumes contained by these redshifts, we extrapolate our
   nearby local sample of galaxies and their predicted CCSRs to both the sky coverage areas and the respective redshifts for the
   three surveys, taking into account the SFH of the universe, the
   redshifting of light from distant CCSNe and the
   time-dilation of both the CCSN lightcurves and the CCSRs. Using the CCSN detection distributions obtained from the
   MCSs we then determine the detected CCSR for each of the surveys. The
   results can be found in Table \ref{CCSN_survey_rates}.

\begin{table*}
\caption{\textrm{Predicted CCSRs for PS1, PS4 and LSST}\label{CCSN_survey_rates}}
  \begin{center}
     \begin{tabular}{c c c c c}
     \hline\hline
     & \multicolumn{4}{c}{\bf Detected Core-Collapse SN Rates (SNe yr$^{-1}$)}\\
     \hline
     {\bf 12+log(O/H)} & {\bf DR5 Sample ($z<0.033$)} & {\bf PS1} & {\bf PS4} & {\bf LSST}\\
     \hline
     {\bf No Limit} & 103.5 & 24\,095 & 68\,620 & 160\,249\\
     {\bf $<8.4$} & 9.8 & 3\,046 & 8\,343 & 19\,627\\
     {\bf $<8.3$} & 5.3 & 1\,640 & 4\,491 & 10\,566\\
     {\bf $<8.2$} & 2.3 & 786 & 2\,153 & 5\,064\\
     {\bf $<8.1$} & 0.9 & 278 & 762 & 1\,794\\
     \hline
     \end{tabular}
   \end{center}
\end{table*}

\section{Discussion and comparison with other surveys}
\label{sect:disc}

Having considered three different survey strategies designed to search
for CCSNe, specifically in low-metallicity environments (a
pointed survey of catalogued low-metallicity galaxies 
using both a single and a network of
2.0m telescopes, a volume-limited survey using the Pan-STARRS all-sky
surveys and finally a magnitdue limited survey using future all-sky
surveys), it is now appropriate to compare these different
strategies, the numbers of CCSNe we expect to detect and the various
limitations of each strategy.

Using a single 2.0m telescope or a network of seven 2.0m telescopes to
perform a pointed survey of low-metallicty galaxies
within $z=0.04$ we expect to
detect approximately \mbox{1.3 CCSNe yr$^{-1}$} and \mbox{9.3 CCSNe
yr$^{-1}$} respectively in host galaxies of very low-metallicity,
exhibiting oxygen abundances \mbox{$12+{\rm \log(O/H)} < 8.2$}. When
considering the relative numbers of similar events that we would hope
to detect using a volume-limited
survey with PS1 we estimate that we
would detect approximately \mbox{12.5 CCSNe yr$^{-1}$} and, extending
the search to $z=0.056$, PS4 would detect \mbox{18.3 CCSNe
yr$^{-1}$}. A drawback of the pointed low-metallicity survey is the
fact that $\sim$1\,000 hrs of telescope time would be required of the
network of seven 2.0m telescopes. Unlike the network of 2.0m
telescopes, the all-sky surveys have the additional
advantage of not
being limited by the number of galaxies that they can feasibly observe
and as a result shall be capable of observing CCSN events in all
environments and not only those in the lowest-metallicity host
galaxies, helping us to reduce the possibility of missing any rare
CCSN events of extreme interest and also gaining the capability of
determining accurate relative rates of SNe within a large but limited
volume, expecting to detect a total of \mbox{$\sim$570 CCSNe
yr$^{-1}$} with PS1 and \mbox{$\sim$1\,300 CCSNe yr$^{-1}$} with PS4.

To estimate the amount of 8.0m telescope time that we would require to
spectroscopically follow the $\sim$13 CCSNe found with the PS1 volume-limited survey in low-metallicity environments \mbox{($12+{\rm
\log(O/H)} < 8.2$)}, we consider a CCSN occurring at the average
distance of the survey, that is \mbox{$\sim$120 Mpc}. Depending on how
early we catch this SN, we hope to observe it at various epochs during
its evolution and into its tail phase. Suggesting an average absolute
magnitude of -15 for these epochs equates as an observed apparent
magnitude of 20.7 at 120 Mpc, including a typical galactic extinction
of 0.3 magnitudes. To determine the time required from a typical 8.0m
telescope for spectroscopic follow-up we use the VLT FORS2 exposure
time calculator, selecting a suitable grism and requiring a
signal-to-noise of 50 per spectral element for each spectrum. To reach
this signal-to-noise level we would require a $\sim$3\,000 sec on
source integration. If we were to spectroscopically follow the 13
events, requiring 5 epochs of observations per SN, we would then
require $\sim$70 hrs of 8.0m telescope time per year.

An obstacle that would hinder both the pointed survey of
low-metallicity galaxies and the volume-limited Pan-STARRS surveys is
the incompleteness of any galaxy catalogue that could presently be
produced.  In the former case one would obviously need an all-sky
target list for a pointed survey as we have demonstrated that the SDSS
footprint does not have enough low-metallicity galaxies to make a
search particularly fruitful (only 1.3 CCSNe\,yr$^{-1}$ is likely to
be found). In the case of the wide area search of Pan-STARRS one would
ideally like to quantify all potential low-metallicity systems within
a local volume of the Universe so that new discoveries can be quickly
cross-matched and the optical transients in these galaxies immediately
selected for follow-up.  Hence it would be extremely useful to have an
SDSS DR5 type catalogue over the entire sky.  As previously discussed
(see \ref{sec_f2}) it is currently possible to compile a catalogue of
$\sim$ $10^5$ galaxies within the sky area covered by Pan-STARRS and
within the redshift range $z<0.04$.  This catalogue is far from
complete, however when PS1 completes a full cycle of sky observations
we would have a deep photometric catalogue with which to search for
CCSNe and classify the host galaxies.

It will be challenging to identify and select genuine CCSNe in the
magnitude-limited, all sky-surveys.  Without a defined volume limit it
becomes impossible to describe accurate relative rates of SNe and
without the prior knowledge of the SN hosts it will be very difficult
to segregate CCSNe from the large numbers of other transitory events,
especially more distant SNe Type Ia in apparently faint (but perhaps
intrinsically bright) hosts. Most of the SNe are likely to be too
distant to have any catalogued host and one of the challenges is to
classify objects from the information gathered in the all-sky surveys
themselves e.g. lightcurve matching, identification and photometric
redshift measurement of the host, estimated energy of the transient.
If it is possible to overcome these challenges the numbers of detected
CCSNe expected from these unbiased, all-sky and magnitude-limited
surveys shall overwhelm those numbers from the other survey straegies
far above an order of magnitude (see Table \ref{CCSN_survey_rates}).

A pioneering survey in this area has been the Texas Supernova Search
(TSS), which has now evolved into the ROTSE Supernova Verification
Project (RSVP)
\citep{2005AAS...20717102Q, 2007AAS...21110505Y}.  The Robotic Optical
Transient Search Experiment (ROTSE) presently consists of four 0.45m
robotic telescope located around the globe, each with a 1.85 square
degree field of view, dedicated to detecting and observing optical
transients, with an emphasis on GRBs. The TSS has used the ROTSE-IIIb
(MacDonald Observatory Texas) to perform a
wide-field search for nearby SNe. 
The RSVP uses all four ROTSE-III telescopes
to discover SNe and other transient variables, by imaging areas of
the sky that host nearby galaxy clusters. The typical cadence is 1-3
days with a typical limiting magnitude of $\sim$18.5 mags. To date
the ROSTE-IIIb telescope has discovered 50 SNe in total at mean redshift
of $z=0.049$; 34 Type Ia and 16 Type II (5 of which have been
classified as Ic and 5 as IIn). The most remarkable result of this
small scale SN survey is that they have claim over the discovery of
the three most luminous SN ever detected; SN2006tf (-20.7 mag),
SN2006gy (-22 mag) and SN2005ap (-22.7 mag). SN 2006tf and SN 2006gy
were both classified as a Type IIn \citep{2007CBET..793....1Q,
2006CBET..644....1Q} and possibly had Luminous Blue Variable (LBV)
star progenitors with explosions speculated to have been triggered by
pulsation pair instability \citep{2008arXiv0804.0042S,
2006astro.ph.12617S,2007Natur.450..390W}. SN 2005ap was classified as
a Type II \citep{2005CBET..116....1Q} and exhibited a broad H$\alpha$
P-Cygni profile. The explosion of the progenitor, which still had its
Hydrogen envelope intact, was possibly powered by either a GRB type
engine or a pair-instability eruption
\citep{2007ApJ...668L..99Q}. Although these type of super-luminous SNe
may be intrinsically rare, it is more likely that they have remained
undiscovered in the past due to the fact that they have been missed or
misclassified as AGN in pointed surveys as they have occurred close to
the cores of bright galaxies. The image-subtraction techniques used by
area surveys, such as TSS and RSVP, will ensure that these types of
interesting events will continue to be discovered. 
Taking a spectrum of the SN Type IIn 1998S
at peak (Fassia et al. 2000) and redshifting it to estimate the
$K$-corrections that SNe similar to 2006gy would require at various
redshifts, we predict that we would detect these events to $z\sim 0.6$
for PS1, $z\sim 0.8$ for PS4 and $z\sim 0.9$ for LSST. For PS4 and
LSST these limits correspond to a search volume $\sim 2.9$ and $\sim
4.4$ times greater than PS1.

\begin{figure*}[b]
  \begin{center}
  \includegraphics[totalheight=0.15\textheight]{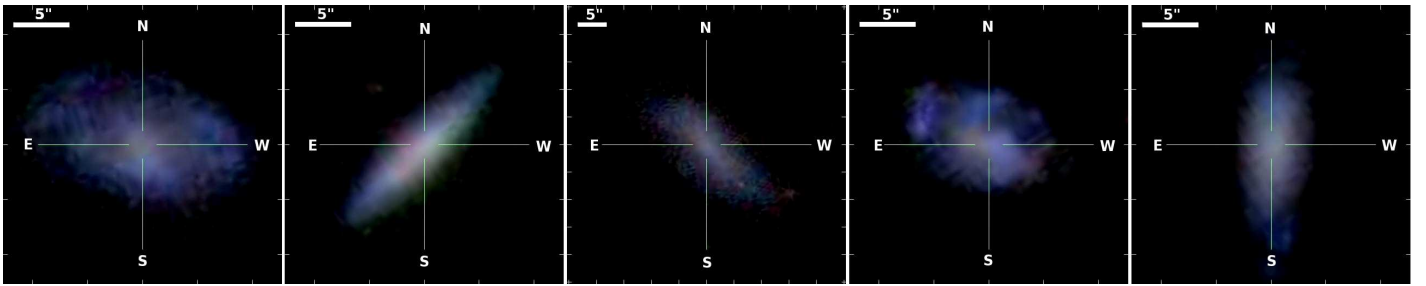}
  \caption{\textrm{Of the 12 SNe discovered by the Nearby Supernova Factory that were hosted by SFGs in the SDSS DR5 catalogue, 5 were hosted by galaxies with 12+log(O/H) $<$ 8.4. The host shown are SDSS J140737.23+384047 (SN 2007fg), SDSS J160205.11+294338 (SN 2007fz), SDSS J135042.68+400208 (SNF20080324-010), SDSS J154404.32+275335 (SNF20080515-004) and SDSS J150031.40+552210 (SNF20080614-002).}\label{NSF-gal}}
  \end{center}
\end{figure*}

The Nearby Supernova Factory (NSF) uses ultra-wide
field CCD mosaic images from the Near-Earth Asteriod Tracking (NEAT) and Palomar-Quest survey
programs with the aim of finding low-redshift SNe Ia
\citep{2002SPIE.4836...61A}. Within the volume-limited ($z<0.04$) search with PS1 we predict relative-rates of SNe II, SNe Ib/c and SNe IIn as 57.8\%, 38.2\% and 4.0\% and at mean redshifts of 0.0318, 0.0316 and 0.0321 respectively. In its lifetime the NSF had discovered and classified 85 CCSNe within $z<0.04$ with relative-rates of SNe II, SNe Ib/c and SNe IIn of 67.1\%, 20.0\% and 12.9\% at mean redshifts of 0.0283, 0.0292 and 0.0284 respectively. The reason that the relative-rates of CCSNe types predicted by our PS1 volume-limited search and those discovered by the NSF are somewhat discrepant may be due in part to the photometric screening of SN candidates that the NSF perform prior to spectroscopically classifying candidates. The NSF aim to find young SNe Ia that are nearby yet are within the Hubble-flow and therefore seem to classify SN candidates that are on the rise and that are close to the limiting-magnitude of their search (R $\sim$ 20 mags), as can be seen from the relative-magnitude distributions of classified SNe in Table \ref{NSF_PS1}.  However it could be argued that because of the high-metallicity bias of many current and historic SN searches, the relative-rates of nearby CCSNe compiled by  \cite{smartt_et_al}, and used in our simulations, are also biased toward higher metallicity. A high-metallicity bias in these relative-rates would be seen as an over-estimation of the true SN Ib/c to SN II ratio; as seen when compared to the relative-rates discovered by the NSF. Further evidence supporting this hypothesis results from cross-matching NSF discovered SNe with our catalogue of SDSS DR5 SFGs within $z<0.04$. A total of 12 CCSNe are matched with SDSS galaxies, 5 of which occurring in galaxies  with 12+log(O/H)$<$8.4 (see Fig. \ref{NSF-gal}). This high fraction of CCSNe discovered in low-metallicity environments may again be due to selection criteria of the NSF (possibly looking for SNe occurring in apparently faint hosts) but remains encouraging for our plans to search for low-metallicity CCSNe with future all-sky surveys.

\begin{table}
\caption{\textrm{Comparison between the relative magnitudes of CCSNe we predict with a volume-limited ($z<0.04$) search with PS1 and those discovered within the same volume by the Nearby Supernova Factory.}\label{NSF_PS1}}
  \begin{center}
    \begin{tabular}{c c c}
      \hline\hline
      {\bf Magnitude Range} & {\bf PS1 Predictions}&{\bf Classified by NSF}\\
      \hline
      $< 16$ & 3.5 \% & 3.5 \%\\
     16 - 17 & 6.1 \% & 5.9 \%\\
     17 - 18 & 12.6 \% & 3.5 \%\\
      18 - 19 & 20.3 \%& 47.1 \%\\
     19 - 20 & 25.4 \% & 40.0 \%\\
     $>20$ & 32.2 \% & 0.0 \%\\
    \end{tabular}  
  \end{center}
\end{table}

Another future all-sky survey that has potential for SNe discoveries is 
Gaia : the European Space Agency's `super-Hipparcos' satellite with
the objective of creating the most precise three-dimensional map of
the Galaxy \citep{2005ASPC..338....3P}. The satellite shall have many
other additional capabilities including the ability to detect nearby
SNe (within a few hundred Mpc), and predicted for launch in December
2011 it is a potential competitor of Pan-STARRS and
LSST. \cite{2003MNRAS.341..569B} have performed a feasibility study
similar to this study for Gaia and have predicted that the satellite
shall detect \mbox{$\sim$1\,420 CCSNe yr$^{-1}$} using a magnitude-limited
survey strategy. Capable of detecting objects brighter than 20th
magnitude, 1.5 magnitudes brighter than the $g$=21.5 limit for PS1,
Gaia has the ability to survey a volume of only an eighth of the depth
that PS1 shall survey. \citet{2003MNRAS.341..569B} employed a galaxy
catalogue which more than likely neglected low-luminosity galaxies,
which would result in the CCSN rate being under-predicted by a factor
of up to $\sim$2 fewer SNe. Hence if we scale the
\citeauthor{2003MNRAS.341..569B} numbers by $\sim$16, the final
numbers should be comparable with our PS1 estimates, predicting
\mbox{$\sim$22\,720 CCSNe yr$^{-1}$}, which is very close to the rate
that we have predicted for PS1 using our MCSs, \mbox{$\sim$24\,000
CCSNe yr$^{-1}$} (see Table \ref{CCSN_survey_rates}). However, we note
that these numbers are likely somewhat over optimistic due to the
treatment of host galaxy extinction in our MCS.

\section{Conclusions}
Having determined oxygen abundances, star-formation rates and CCSN rates for all
spectroscopically typed star-forming galaxies in the Sloan Digital Sky
Survey Data Release 5 within $z=0.04$, we have used Monte-Carlo simulations to predict the fraction of
these CCSNe that we can expect to detect using different survey
strategies. Using a single 2m telescope (with a standard CCD camera) 
search we predict a detection rate of
$\sim$1.3 CCSNe yr$^{-1}$ in galaxies with metallicities below
$12+\log({\rm O/H})<8.2$ which are within a volume that will allow
detailed follow-up with 4m and 8m telescopes ($z=0.04$). With a
network of seven 2m telescopes we estimate $\sim$9.3 CCSNe yr$^{-1}$ 
could be found, although this would require more than 
1000\,hrs of telescope time allocated to the network. 
Within the same radial distance, a 
volume-limited search with the future Pan-STARRS
PS1 all-sky survey should uncover 12.5 CCSNe yr$^{-1}$  in low metallicity
galaxies. Over a period of a a few years this would allow a detailed comparison of their properties. We have also extended our 
calculations to determine the total numbers of CCSNe that can potentially be 
found in magnitude-limited surveys 
with PS1 (24\,000 yr$^{-1}$, within $z \lesssim 0.6$), 
PS4 (69\,000 yr$^{-1}$, within $z \lesssim 0.8$ ) 
and LSST (160\,000 yr$^{-1}$, within $z \lesssim0.9 $) surveys.

All considered, a final strategy chosen to searching for CCSNe in low-metallicity environments shall realistically involve both a volume-limited and a magnitude-limited all-sky survey in order to include a volume-limited galaxy sample with which to accurately determine relative SN rates and have some prior knowledge of the host galaxy characteristics, but yet not exclude the potential of detecting rare CCSN events that would have otherwise been missed had we only considered the volume-limited survey strategy. 

With the huge number of CCSNe predicted to be detected, these all-sky surveys are set to serve as a catalyst concerning our understanding of CCSNe; including their varying characteristics with metallicity, the relative rates of the various types of SNe and of extremely rare events similar to SN 2006jc, SN 2006gy and more than likely events the nature of which have not yet been observed.

\begin{acknowledgements}

    Funding for the Sloan Digital Sky Survey (SDSS) and SDSS-II has been provided by the Alfred P. Sloan Foundation, the Participating Institutions, the National Science Foundation, the U.S. Department of Energy, the National Aeronautics and Space Administration, the Japanese Monbukagakusho, and the Max Planck Society, and the Higher Education Funding Council for England. The SDSS Web site is http://www.sdss.org/.

    The SDSS is managed by the Astrophysical Research Consortium (ARC) for the Participating Institutions. The Participating Institutions are the American Museum of Natural History, Astrophysical Institute Potsdam, University of Basel, University of Cambridge, Case Western Reserve University, The University of Chicago, Drexel University, Fermilab, the Institute for Advanced Study, the Japan Participation Group, The Johns Hopkins University, the Joint Institute for Nuclear Astrophysics, the Kavli Institute for Particle Astrophysics and Cosmology, the Korean Scientist Group, the Chinese Academy of Sciences (LAMOST), Los Alamos National Laboratory, the Max-Planck-Institute for Astronomy (MPIA), the Max-Planck-Institute for Astrophysics (MPA), New Mexico State University, Ohio State University, University of Pittsburgh, University of Portsmouth, Princeton University, the United States Naval Observatory, and the University of Washington.
    
  This research has made use of the CfA Supernova Archive, which is funded in part by the National Science Foundation through grant AST 0606772.
  
  This work, conducted as part of the award "Understanding the lives of
massive stars from birth to supernovae" (S.J. Smartt) made under the
European Heads of Research Councils and European Science Foundation
EURYI (European Young Investigator) Awards scheme, was supported by
funds from the Participating Organisations of EURYI and the EC Sixth
Framework Programme. SJS and DRY thank the Leverhulme Trust and
DEL for funding. SM acknowledges financial support from the Academy of Finland, project 8120503.

\end{acknowledgements}

\bibliography{bibtex/mybib}

\end{document}